# Two-sorted algebraic decompositions of Brookes's shared-state denotational semantics

Yotam Dvir[1], Ohad Kammar[2], Ori Lahav[1], and Gordon Plotkin[2]

[1] Tel Aviv University yotamdvir@mail.tau.ac.il orilahav@tau.ac.il
[2] University of Edinburgh ohad.kammar@ed.ac.uk gdp@inf.ed.ac.uk

**Abstract.** We define a two sorted equational theory of algebraic effects that models concurrent shared state with preemptive interleaving, recovering Brookes's seminal 1996 trace-based model precisely. The decomposition allows us to analyse Brookes's model algebraically in terms of separate but interacting components. The multiple sorts partition terms into layers. We use two sorts: a "hold" sort for layers that disallow interleaving of environment memory accesses, analogous to holding a global lock on the memory; and a "cede" sort for the opposite. The algebraic signature comprises of independent interlocking components: two new operators that switch between these sorts, delimiting the atomic layers, thought of as acquiring and releasing the global lock; non-deterministic choice; and state-accessing operators. The axioms similarly divide cleanly: the delimiters behave as a closure pair; all operators are strict, and distribute over non-empty non-deterministic choice; and non-deterministic global state obeys Plotkin and Power's presentation of global state. Our representation theorem expresses the free algebras over a two-sorted family of variables as sets of traces with suitable closure conditions. When the held sort has no variables, we recover Brookes's trace semantics. We define several other single- and two-sorted theories to elucidate the connection to Brookes's model via translation embeddings and equivalences.

## 1 Introduction

We decompose Brookes's pioneering denotational model of concurrent shared state under preemptive interleaving [7] using algebraic effects [35]. This model possesses several desirable features in the area of denotational models for programming languages with concurrent features. (I) It is based on traces, an elementary sequential gadget. (II) It is fully compositional, as in traditional denotational semantics for shared-state [e.g. 15, 17]. Each syntactic programming construct, including parallel composition, has a corresponding semantic operation combining the meanings of its constituents. Such full compositionality contrasts with some recent models in this area that require additional 'semantic post-processing': some form of quotient, pruning of auxiliary mathematical constructs, reasoning up-to behavioural equivalence; or capture only sequential blocks, reasoning about the parallel composition on a separate layer [e.g. 8, 9, 20, 25]. (III) Subsequent variations and extensions [5, 44, 45], as well as adaptations to relaxed memory models [13, 14, 19], attest to its versatility, making it a cornerstone in the denotational semantics for concurrent languages with side-effects. (IV) It achieves a high level of abstraction, evident in the many compiler transformations that the model supports, including the most common memory access introductions and eliminations, and the laws of parallel programming. Moreover, Brookes showed the model to be fully abstract in a language extended with the `await` construct, which blocks execution until all memory locations contain a given tuple of values, and then atomically updates them to contain another tuple of values. This construct is not a natural programming construct, but is clearly suggested by Brookes's semantics.

Plotkin and Power's modern theory of *algebraic effects* [35] refines Moggi's monadic approach [30] with algebraic theories. The algebraic approach informs the monadic structure by identifying semantic counterparts to syntactic constructs and axiomatising their semantics equationally. The monadic structure emerges through the well-established connection between algebraic theories and monads [27] via *representation theorems*. For example: global state emerges by axiomatising memory lookup and update [35] and a representation theorem involving the state monad; non-determinism emerges by axiomatising semi-lattices and a representation theorem involving the powerdomains [15, 32]; and so on. The algebraic perspective may offer insights into the making of the denotational semantics. It can suggest methods for combining different effects and modularly augment a semantics with a given computational effect [17].

The connection between algebraic effects and concurrency has long been emphasised. For example, the ability to use algebraic effects, without any axioms, and their *effect handlers* [4, 37, 38] to allow users to define their own schedulers was the original motivation for their implementation in the OCaml programming language [10, 11, 40]. Nonetheless, exhibiting abstract models such as Brookes's algebraically via equational axiomatisation of syntactic constructs has proved challenging. Our own previous algebraic model [12] invalidates a key transformation, reflecting a fundamental limitation.

To overcome this limitation, we use multi-sorted algebraic theories, a direction that was raised in personal discussions since the earliest work on algebraic effects [35]. A multi-sorted algebraic term decomposes into layers. Our two sorts represent two modes of interaction between a program fragment and its concurrent environment. A "hold" sort ($\bullet$) provides a reasoning layer in which the environment may not interfere, whereas in the "cede" sort ($\circ$) it may. We provide two operators that switch between these sorts. Our core idea is to axiomatise these operators as a *closure pair*, an established order-theoretic special Galois-connection, the dual to the domain-theoretic embedding-projection pairs [2]. Additionally, we axiomatise strict distributivity of the closure pair over non-determinism. The remaining axioms, all in the hold sort, are strikingly independent from these axioms. In our shared-state theory $\mathbb{S}$, the remaining axioms are precisely those of non-deterministic global state.

We prove, twice over, that $\mathbb{S}$ recovers Brookes's model in the cede sort. First, using sets of traces akin to Brookes's, we define a representation of $\mathbb{S}$. The representation recovers Brookes's model via the adjunction that forgets the hold sort. Second, we define three algebraic theories for Brookes's `await` and its sequential variant, relating them to global-state, shared-state, and each other via translation embeddings and equivalences. The theory for concurrent `await` is straightforwardly represented by Brookes's model, and embeds in $\mathbb{S}$'s cede sort.

*Caveats* In our development, we opt for mathematical simplicity whenever possible. For example, we use countable-join semilattices instead of finite-join semilattices to represent non-determinism. This choice streamlines the development leading up to the representation theorem, allowing us to use countable sets instead of finitely generated ones. We also do not treat recursion to avoid the complexity that a domain-theoretic account incurs. The resulting model—identical to Brookes's—coincides with the elided domain-theoretic model over discrete predomains. This model also supports iteration (i.e. `while`-loops) without change thanks to countable-joins. It also supports first-order recursion without change by equipping it with a domain-theoretic structure. These compromises let us focus on the core concepts, and provide a relatively elementary exposition and a clear presentation of the underlying idea, motivating future inquiry.

## 2 Overview

Equational theories study terms constructed from algebraic operators (§3.1). In Plotkin and Power's algebraic theory of effects, the operators represent fundamental program effects, and their arguments represent continuations. Equational axioms reflect fundamental relationships between the operators. The equations that hold in the theory, reflecting the semantics as a whole, are those that follow from its axiomatic presentation by equational logic (§3.2).

In the global-state theory, a theory for sequential stateful computation, the operators $\mathsf{U}$ and $\mathsf{L}$ represent updating and looking up bits in memory. For example, consider the global-state term $\mathsf{U}_{\mathtt{y},\mathtt{0}}\,\mathsf{L}_{\mathtt{y}}(3, \mathsf{U}_{\mathtt{x},\mathtt{1}}\,\mathsf{U}_{\mathtt{y},\mathtt{1}}\,7)$. After updating `y` to `0`, the computation looks `y` up: if it finds `0`, it returns 3; if it finds `1`, it updates `x` and `y` to `1` in succession, then returns 7. Between the update and the lookup, the value at `y` cannot change. Therefore, the computation finds the value `0`, and takes the left-hand continuation. The global-state axiom (UL) $\mathsf{U}_{\ell,b}\,\mathsf{L}_{\ell}(x_{\mathtt{0}}, x_{\mathtt{1}}) = \mathsf{U}_{\ell,b}\,x_b$ reflects this fact. By (UL), $\mathsf{U}_{\mathtt{y},\mathtt{0}}\,\mathsf{L}_{\mathtt{y}}(3, \mathsf{U}_{\mathtt{x},\mathtt{1}}\,\mathsf{U}_{\mathtt{y},\mathtt{1}}\,7) = \mathsf{U}_{\mathtt{y},\mathtt{0}}\,3$ holds in global state.

A representing monadic model for an equational theory (§3.3) interprets the algebraic operators as corresponding operations over the model's domain; such that each term, up-to equality in the theory, is represented uniquely in the domain. Interpretations respect the theory; that is, applying an operation to representations of terms results in the representation of the corresponding operator applied to said terms: $[\![O]\!]_{\mathrm{op}}([\![t_1]\!]_{\mathrm{term}}, \dots, [\![t_\alpha]\!]_{\mathrm{term}}) = [\![O(t_1, \dots, t_\alpha)]\!]_{\mathrm{term}}$.

For example, global-state terms are represented by memory-manipulating functions in the state-monad model. This model interprets update by precomposing a state update $[\![\mathsf{U}_{\ell,b}]\!]_{\mathrm{op}} f = f \circ [\ell \mapsto b]$; and interprets lookup by passing the input memory state $\sigma$ along to the $\sigma_\ell$-continuation $[\![\mathsf{L}_\ell]\!]_{\mathrm{op}}(f_0, f_1) = \lambda\sigma.\, f_{\sigma_\ell}\sigma$. In this way, global state recovers the (historically precedent) state monad.

The state monad does not account for concurrent interference. The monad underlying Brookes's denotational semantics does, by using sequences of transitions to denote potential behaviours (§6.1). Each transition $\langle\sigma, \rho\rangle$ in sequence means that the computation, by relying on exclusive access to the memory at state $\sigma$, can guarantee to provide the state $\rho$ and yield to the environment.

Following the tradition of algebraic effects, we wish to recover Brookes's model using an equational theory for shared state. This is rather straightforward using a single-sorted theory $\mathsf{B}$ (§6.3) in which transitions appear as operators. However, this theory is dissatisfying, for two reasons. (I) Transitions do not correspond to familiar programming constructs, but to Brookes artificial `await` construct. (II) Conceptualising shared state as global state with concurrent interference, we expect the global-state effects to be present in a theory of shared state, and the equations between them to hold when interference is prohibited.

A more appropriate approach adds an operator $\mathsf{Y}$ to global state for yielding control to the environment. Otherwise, the computation has exclusive access to memory. This direction lead to a Brookes-like model [12]. However, unlike Brookes's model, it does not validate the Irrelevant Read Introduction (IRI) program transformation, which introduces a read instruction that possibly yields to the environment and discards the value it read. The IRI transformation is useful as a stepping stone to other practical transformations, such as common-subexpression elimination from conditionals, and consequently loops. Invalidating IRI seems to be a fundamental limitation of the yield-operator approach, as we show in the appendix (§A). In retrospect, we can pinpoint the issue to $\mathsf{Y}$ both releasing exclusive access to memory, and acquiring it back. Our key insight is that the mode of computation that cedes access to memory needs to be explicit, decomposing $\mathsf{Y}$ into a pair of mode-switching operators.

The remaining structure falls into place straightforwardly and naturally, in our proposed two-sorted theory for shared state $\mathbb{S}$ (§4). Each sort represents a computation mode: *hold* ($\bullet$) represents the computation's exclusive access to memory; *cede* ($\circ$) represents positions in which the environment may interfere. Each operator has a sort and expects each continuation to have a specific sort. In $\mathbb{S}$, update $\mathsf{U} : \bullet\langle\bullet\rangle$ and lookup $\mathsf{L} : \bullet\langle\bullet, \bullet\rangle$ are $\bullet$-sorted and expect $\bullet$-sorted continuations, allowing us to reason about interference-free stateful interactions.

The theory $\mathbb{S}$ also supports non-deterministic choice in both sorts. For example, the term $(\mathsf{U}_{\mathtt{x},\mathtt{1}}\, 2 \vee \mathsf{U}_{\mathtt{x},\mathtt{1}}\, 5)$ either updates `x` to `1` and returns 2, or updates `x` to `1` and returns 5. We axiomatise $\mathbb{S}$ such that each operator distributes over non-deterministic choice, e.g. $(\mathsf{U}_{\mathtt{x},\mathtt{1}}\, 2 \vee \mathsf{U}_{\mathtt{x},\mathtt{1}}\, 5) = \mathsf{U}_{\mathtt{x},\mathtt{1}}(2 \vee 5)$. We order terms by potential behaviours $(l \geq r) \coloneqq (l = l \vee r)$, a partial-order for $\mathbb{S}$-equality (§3.2).

The mode-switching operators of $\mathbb{S}$ are $\lhd : \circ\langle\bullet\rangle$ and $\rhd : \bullet\langle\circ\rangle$. That is, $\lhd$ is $\circ$-sorted and expects a $\bullet$-sorted continuation, and vice versa for $\rhd$. We think of them as delimiting atomic blocks, or acquiring and releasing an abstract global lock. We axiomatise them in $\mathbb{S}$ by strict distributivity over countable joins, i.e. (ND-$\lhd$) $\bigvee_{i<\alpha} \lhd x_i = \lhd \bigvee_{i<\alpha} x_i$ and (ND-$\rhd$) $\bigvee_{i<\alpha} \rhd x_i = \rhd \bigvee_{i<\alpha} x_i$; and:

**Empty ($\lhd \rhd y = y$).** An empty atomic block has no observable effect.
**Fuse ($\rhd \lhd x \geq x$).** Fusing atomic blocks eliminates potential interference.

These axiomatise $\lhd$ and $\rhd$ as an *(insertion)-closure pair* [e.g. 2].

Each $\mathbb{S}$-term denotes a set of sort-delimited traces (§5.1), which generalise Brookes's traces. For example, $t \coloneqq \mathsf{U}_{\mathtt{y},\mathtt{0}} \rhd \lhd \mathsf{L}_{\mathtt{y}}(3, \mathsf{U}_{\mathtt{x},\mathtt{1}}\, \mathsf{U}_{\mathtt{y},\mathtt{1}} \rhd 7)$ denotes a set that includes $\bullet\langle\left(\begin{smallmatrix}\mathtt{x}\mapsto\mathtt{1}\\ \mathtt{y}\mapsto\mathtt{1}\end{smallmatrix}\right), \left(\begin{smallmatrix}\mathtt{x}\mapsto\mathtt{1}\\ \mathtt{y}\mapsto\mathtt{0}\end{smallmatrix}\right)\rangle\langle\left(\begin{smallmatrix}\mathtt{x}\mapsto\mathtt{1}\\ \mathtt{y}\mapsto\mathtt{1}\end{smallmatrix}\right), \left(\begin{smallmatrix}\mathtt{x}\mapsto\mathtt{1}\\ \mathtt{y}\mapsto\mathtt{1}\end{smallmatrix}\right)\rangle\circ 7$. Indeed, we can read off a corresponding computation from $t$, initially holding the lock in the state $\left(\begin{smallmatrix}\mathtt{x}\mapsto\mathtt{1}\\ \mathtt{y}\mapsto\mathtt{1}\end{smallmatrix}\right)$ of both bits `1`: the computation updates `y` to `0`, then yields to the environment before looking `y` up; finding `1`, it updates `x` and `y` to `1`, then releases the lock and returns 7. Brookes's original traces correspond to those delimited by $\circ$ on both ends.

With these traces we define our two-sorted generalisation of Brookes's model, and prove that it represents $\mathbb{S}$ (§5.2). The $\circ$-sorted $\circ$-valued fragment (§6.2), which represents the "block-closed" terms, is Brookes's original model (§6.1).

We also provide an algebraic perspective on the representation, by translation embedding (§6.4) the transitions-theory $\mathsf{B}$ in $\mathbb{S}$'s $\circ$ sort (§6.5). It translates $\langle\sigma, \rho\rangle$ into $\lhd \{\!|\sigma, \rho|\!\} \rhd$, where $\{\!|\sigma, \rho|\!\} : \bullet\langle\bullet\rangle$ is given by global-state operators (§5.2).

## 3 Preliminaries

We present a standard treatment of countably-infinitary multi-sorted equational theories and their free models [e.g. 3, 43], straightforwardly generalising the single-sorted case by assigning sorts to functions and their arguments. The reader may choose to skim/skip this section, consulting it as necessary.

### 3.1 Terms

We define the logical language of multi-sorted equational logic. The basic vocabulary of multi-sorted algebra is parameterised by a set **sort** whose elements $\Box, \Diamond$ we call *sorts*. We will mostly focus on the *single-sorted* case ($\mathbf{sort} = \{\star\}$) and the *two-sorted* case ($\mathbf{sort} = \{\bullet, \circ\}$). A **sort**-*scheme* $\vec{\Box} \in \operatorname{Scheme} \mathbf{sort}$ is a countable sequence of sorts from **sort**, i.e. a finite sequence $\vec{\Box} = \langle \Box_0, \ldots, \Box_{n-1} \rangle$ of length $n$, or countably infinite sequence $\vec{\Box} = \langle \Box_0, \Box_1, \ldots \rangle$ of length $\omega$, where $\Box_i \in \mathbf{sort}$ for all $i$. For example: the empty scheme $\mathbf{0} \coloneqq \langle\rangle$ of length 0; and the constant schemes $\alpha \cdot \Box \coloneqq \langle \Box \rangle_{i<\alpha}$ of length $\alpha$. We write $\Box$ for the scheme $1 \cdot \Box$.

A **sort**-*sorted signature* $\Sigma = \langle \mathbf{op}_\Sigma, \mathbf{ar}_\Sigma \rangle$ consists of a set of *operators* $\mathbf{op}_\Sigma$ and an *arity* assignment $\mathbf{ar}_\Sigma : \mathbf{op}_\Sigma \to \mathbf{sort} \times \operatorname{Scheme} \mathbf{sort}$. For $O \in \mathbf{op}_\Sigma$ with $\mathbf{ar}_\Sigma\, O = \left\langle \Box, \langle \Diamond_i \rangle_i \right\rangle$, we write $(O : \Box\, \langle \Diamond_i \rangle_{i<\alpha}) \in \Sigma$. The operator $O$ will allow us to construct a $\Box$-sort term with a tuple of terms, with the $i^{\text{th}}$ subterm having sort $\Diamond_i$. For single-sorted arities ($\mathbf{sort} = \{\star\}$), we write $O : \alpha$ for $O : \star\, (\alpha \cdot \star)$. A *signature* is a set $\mathbf{sort}_\Sigma$ and a $\mathbf{sort}_\Sigma$-sorted signature we also denote by $\Sigma$.

We will use the following signature to model non-deterministic choice:

*Example 1.* The *join semilattice* single-sorted signature $\mathtt{J}$ consists of two operators: *join* $\vee : 2$, i.e. $\vee : \star \langle \star, \star \rangle$; and *bottom* $\bot : 0$ , i.e. $\bot : \star \langle \rangle$. $\square$

To simplify the formulation of our representation theorem later, we generalise the signature to countable non-deterministic choice operators:

*Example 2.* The *countable-join semilattice* single-sorted signature $\mathtt{V}$ consists of an $\alpha$-ary *choice* operator $\bigvee_{\alpha} : \alpha$ for every $\alpha \leq \omega$. In particular, the signature $\mathtt{J}$ is included with $\alpha = 2$ (join) and $\alpha = 0$ (bottom). $\square$

The final example demonstrates the treatment for multiple sorts:

*Example 3.* The *finite dimensional transformations* signature $\mathtt{M}$ consists of a sort for each pair of natural numbers $\mathbf{sort}_{\mathtt{M}} := \{\mathbf{Hom}\,(m, n) \mid m, n \in \mathbb{N}\}$, an identity operator $\mathrm{Id}_n : \mathbf{Hom}\,(n, n)\,\langle\rangle$ for each $n \in \mathbb{N}$, and, for each triple $m, n, k \in \mathbb{N}$, a composition operator $(\circ_{m,n,k}) : \mathbf{Hom}\,(m, k)\,\langle \mathbf{Hom}\,(n, k)\,, \mathbf{Hom}\,(m, n)\rangle$. $\square$

A signature generates a language of algebraic terms as follows. A **sort**-*family* $\boldsymbol{X} \in \mathbf{Set}^{\mathbf{sort}}$ is an assignment of a set $\boldsymbol{X}_{\square}$, to each sort $\square \in \mathbf{sort}$. We identify $\mathbf{Set}^{\{\star\}} \cong \mathbf{Set}$, and use a set-like notation to specify families, e.g. $\boldsymbol{X} := \{x : \bullet, y, z : \circ\}$ is the two-sorted family $\boldsymbol{X}_{\bullet} := \{x\}$ and $\boldsymbol{X}_{\circ} := \{y, z\}$. We can turn every **sort**-family $\boldsymbol{X}$ into the set $\oint \boldsymbol{X} := \coprod_{\square \in \mathbf{sort}} \boldsymbol{X}_{\square}$ equipped with the injections $\mathrm{in}_{\square} : \boldsymbol{X}_{\square} \to \oint \boldsymbol{X}$. This construction is a special case of the Grothendieck construction, and lets us track the distinction between sets and families.

For a signature $\Sigma$ and $\mathbf{sort}_{\Sigma}$-family $\boldsymbol{X} \in \mathbf{Set}^{\mathbf{sort}_{\Sigma}}$, define the $\mathbf{sort}_{\Sigma}$-family of $\Sigma$-*terms over* $\boldsymbol{X}$: $\mathrm{Term}^{\Sigma}\boldsymbol{X} \in \mathbf{Set}^{\mathbf{sort}_{\Sigma}}$, $\mathrm{Term}^{\Sigma}_{\square}\boldsymbol{X} := \{t \mid \boldsymbol{X} \vdash_{\Sigma} t : \square\}$ inductively:

$$\frac{(x : \square) \in \boldsymbol{X}}{\boldsymbol{X} \vdash_{\Sigma} x : \square} \qquad \frac{(O : \square \,\langle \Diamond_i \rangle_{i<\alpha}) \in \Sigma \qquad \forall i.\, \boldsymbol{X} \vdash_{\Sigma} t_i : \Diamond_i}{\boldsymbol{X} \vdash_{\Sigma} O\,\langle t_i \rangle_{i<\alpha} : \square}$$

Here, the elements $x \in \boldsymbol{X}_{\square}$, written $(x : \square) \in \boldsymbol{X}$, represent variables of sort $\square$. We may drop the set-brackets left of a trunstile, e.g. $x : \bullet, y, z : \circ \vdash_{\Sigma} y : \circ$; and omit the sorts, especially in the single-sorted case, e.g. $x, y \vdash_{\mathtt{J}} x \vee \bot$. For $t \in \mathrm{Term}^{\Sigma}_{\square}\boldsymbol{X}$, we write $\boldsymbol{X} \vdash_{\Sigma} \psi(...) := t : \square$ to define $\psi$ as $t$, e.g. $x, y \vdash_{\mathtt{J}} \psi(x, y) := x \vee \bot$.

A **sort**-*sorted map* $f : \boldsymbol{X} \to \boldsymbol{Y}$ is a **sort**-indexed tuple of functions between the corresponding sets: $f_{\square} : \boldsymbol{X}_{\square} \to \boldsymbol{Y}_{\square}$, for every $\square \in \mathbf{sort}$. Our development utilises sorted maps extensively. A *(simultaneous) substitution* $\boldsymbol{X} \vdash_{\Sigma} \theta : \boldsymbol{Y}$ is a sorted function $\theta : \boldsymbol{Y} \to \mathrm{Term}^{\Sigma}\boldsymbol{X}$, specifying which $\square$-term $\boldsymbol{X} \vdash_{\Sigma} \theta_{\square} y : \square$ to substitute for each variable $y \in \boldsymbol{Y}_{\square}$. Each such substitution determines a sorted map $[\theta] : \mathrm{Term}\boldsymbol{Y} \to \mathrm{Term}\boldsymbol{X}$ inductively, which we write in post-fix notation:

$$(\boldsymbol{Y} \vdash_{\Sigma} y : \square)\,[\theta] := (\boldsymbol{X} \vdash_{\Sigma} \theta_{\square} y : \square) \qquad (\boldsymbol{Y} \vdash_{\Sigma} O\,\langle t_i \rangle_i)\,[\theta] := (\boldsymbol{X} \vdash_{\Sigma} O\,\langle t_i\,[\theta] \rangle_i)$$

### 3.2 Equational logic

A $\square$-*sorted* $\Sigma$-*equation in context* $\boldsymbol{X}$ is a pair $\langle l, r \rangle \in \mathrm{Term}^{\Sigma}_{\square}\boldsymbol{X}$ of $\square$-sorted $\Sigma$-terms over $\boldsymbol{X}$. We write this situation as $\boldsymbol{X} \vdash_{\Sigma} l = r : \square$, or just $l = r$, and call $l$ the left-hand side (LHS) and $r$ the right-hand side (RHS) of the equation. A *presentation* $\mathfrak{p}$ consists of a signature $\Sigma_{\mathfrak{p}}$ and *axioms*: a set $\mathrm{Ax}_{\mathfrak{p}}$ of $\Sigma$-equations.

$$\frac{\boldsymbol{X} \vdash_{\Sigma_{\mathfrak{p}}} t : \boxbox}{\boldsymbol{X} \vdash_{\mathfrak{p}} t = t : \boxbox} \qquad \frac{\boldsymbol{X} \vdash_{\mathfrak{p}} t_2 = t_1 : \boxbox}{\boldsymbol{X} \vdash_{\mathfrak{p}} t_1 = t_2 : \boxbox} \qquad \frac{\boldsymbol{X} \vdash_{\mathfrak{p}} t_1 = t_2 : \boxbox \qquad \boldsymbol{X} \vdash_{\mathfrak{p}} t_2 = t_3 : \boxbox}{\boldsymbol{X} \vdash_{\mathfrak{p}} t_1 = t_3 : \boxbox}$$

$$\frac{(\boldsymbol{X} \vdash_{\Sigma_{\mathfrak{p}}} t_1 = t_2 : \boxbox) \in \mathrm{Ax}_{\mathfrak{p}}}{\boldsymbol{X} \vdash_{\mathfrak{p}} t_1 = t_2 : \boxbox} \qquad \frac{\boldsymbol{Y} \vdash_{\mathfrak{p}} t_1 = t_2 : \boxbox \qquad \boldsymbol{X} \vdash_{\Sigma_{\mathfrak{p}}} \theta : \boldsymbol{Y}}{\boldsymbol{X} \vdash_{\mathfrak{p}} t_1 [\theta] = t_2 [\theta] : \boxbox}$$

$$\frac{\boldsymbol{Y} \vdash_{\Sigma_{\mathfrak{p}}} t : \boxbox \qquad \boldsymbol{X} \vdash_{\Sigma_{\mathfrak{p}}} \theta, \theta' : \boldsymbol{Y} \qquad \forall (y : \Diamond) \in \boldsymbol{Y}. \boldsymbol{X} \vdash_{\mathfrak{p}} \theta_{\Diamond} y = \theta'_{\Diamond} y : \Diamond}{\boldsymbol{X} \vdash_{\mathfrak{p}} t [\theta] = t [\theta'] : \boxbox}$$

**Fig. 1.** Multi-sorted equational logic with countable arities

*Example 4.* The *join semilattice* presentation $\mathsf{J}$ consists of the signature $\Sigma_{\mathsf{J}} := \mathtt{J}$ of example 1, and the axioms $\mathrm{Ax}_{\mathsf{J}}$ below:

(Associativity) $x \vee (y \vee z) = (x \vee y) \vee z$ (Idempotency) $x \vee x = x$
(Commutativity) $x \vee y = y \vee x$ (Neutrality) $x \vee \bot = x$ $\square$

*Example 5.* The *countable-join semilattice* presentation $\bigvee$ consists of the signature $\Sigma_{\bigvee} := \mathtt{V}$ of example 2, and the axioms $\mathrm{Ax}_{\bigvee}$:

(ND-return) $\bigvee_{i<1} x_i = x_0$
(ND-squash) $\bigvee_{i<\alpha} \bigvee_{j<\beta_i} x_{i,j} = \bigvee_{k<\gamma} x_{fk}$ where $f : \gamma \twoheadrightarrow \coprod_{i<\alpha} \beta_i$ $\square$

*Example 6.* The *finite dimensional transformations* presentation $\mathsf{M}$ consists of the signature $\Sigma_{\mathsf{M}} := \mathtt{M}$ of example 3 and the axioms $\mathrm{Ax}_{\mathsf{M}}$ below, suppressing the sort indices (each axiom scheme includes every possible instantiation):

(L-Id) $\mathrm{Id} \circ f = f$ (R-Id) $f \circ \mathrm{Id} = f$ (Assoc) $f \circ (g \circ h) = (f \circ g) \circ h$ $\square$

Figure 1 presents the deductive system called *equational logic*. We say that a presentation $\mathfrak{p}$ *proves* an equation, writing $\boldsymbol{X} \vdash_{\mathfrak{p}} t_1 = t_2 : \boxbox$, when it is derivable from $\mathrm{Ax}_{\mathfrak{p}}$ using these standard equational reasoning rules, namely: reflexivity, symmetry, transitivity, use of an axiom, substitution, and congruence. This logic is monotone: assuming more axioms allows us to prove more equations. The *algebraic theory* of a presentation $\mathfrak{p}$ is the smallest derivation-closed set of equations containing the axioms. We denote the theory of $\mathfrak{p}$ by $\mathfrak{p}$ as well.

*Example 7.* We can prove $\{x, y : \star\} \vdash_{\mathsf{J}} (x \vee \bot) \vee y = x \vee y : \star$ using an instance of Neutrality and reflexivity with the following instance of congruence:

$$\{z, y : \star\} \vdash_{\mathsf{J}} t := z \vee y : \star \qquad \theta_{\star} := \begin{pmatrix} z \mapsto x \vee \bot \\ y \mapsto y \end{pmatrix} \qquad \theta'_{\star} := \begin{pmatrix} z \mapsto x \\ y \mapsto y \end{pmatrix} \qquad \square$$

When a presentation $\mathfrak{p}$ proves the semi-lattice axioms in one of its sorts $\boxbox$, then the encoding $(\boldsymbol{X} \vdash_{\Sigma_{\mathfrak{p}}} l \leq r : \boxbox) := (\boldsymbol{X} \vdash_{\Sigma_{\mathfrak{p}}} l \vee r = r : \boxbox)$ of inequations as equations in this sort is a preorder that is a partial order w.r.t. $\mathfrak{p}$-equality, i.e. $(\boldsymbol{X} \vdash_{\mathfrak{p}} s \leq t : \boxbox) \wedge (\boldsymbol{X} \vdash_{\mathfrak{p}} t \leq s : \boxbox) \implies (\boldsymbol{X} \vdash_{\mathfrak{p}} s = t : \boxbox)$. We encode $(\geq)$ similarly. Due to the monotonicity property of equational logic, once we have included an axiomatisation of semi-lattices through a subset of the axioms, we may proceed to postulate inequations.

We also use a generalisation of distributivity axioms [18], reproducing familiar arithmetic distributivity equations such as $x \cdot \max\{y_1, y_2\} = \max\{x \cdot y_1, x \cdot y_2\}$, the distributivity of $(\cdot)$ over max in the right-hand side position. We describe the straightforward, but technical generalisation in the appendix (§B). The main message is as follows. In a given presentation $\mathfrak{p}$, if all operators distribute over binary joins in every position, the congruence rule is valid for inequations:

$$\frac{\boldsymbol{Y} \vdash_{\Sigma_{\mathfrak{p}}} t : \Box \qquad \boldsymbol{X} \vdash_{\Sigma_{\mathfrak{p}}} \theta, \theta' : \boldsymbol{Y} \qquad \forall (y : \Diamond) \in \boldsymbol{Y}. \boldsymbol{X} \vdash_{\mathfrak{p}} \theta_{\Diamond} y \leq \theta'_{\Diamond} y : \Diamond}{\boldsymbol{X} \vdash_{\mathfrak{p}} t\,[\theta] \leq t\,[\theta'] : \Box}$$

If a presentation $\mathfrak{p}$ supports semi-lattices in every sort and they distribute over binary joins in every positions, then we say that $\mathfrak{p}$ *supports inequational reasoning*. The theory of $\mathfrak{p}$ then admits Bloom's logic for ordered algebraic theories [6]. We let future work determine the most appropriate variety of inequational logic [34].

Going forward, all of our presentations support inequational reasoning in this sense, and all operators distribute over arbitrary non-empty joins, not just the binary ones. Moreover, they are all strict: $O(\bot, \ldots, \bot) = \bot$ for every operator $(O : \Box \langle \Diamond_i \rangle_{i<\alpha}) \in \Sigma_{\mathfrak{p}}$. Such theories 'absorb' side-effects when their continuations diverge, an inherent 'partial correctness' property of Brookes's model.

### 3.3 Algebras and models

After presenting the proof theory—equational logic—let's turn to the model theory of universal algebra. A $\Sigma$-*algebra* $\mathbf{A}$ consists of a $\mathbf{sort}_\Sigma$-family $\underline{\mathbf{A}} \in \mathbf{Set}^{\mathbf{sort}_\Sigma}$, the *carrier*, and an assignment $\mathbf{A}\,[\![-]\!]_{\mathrm{op}}$, for each operator $(O : \Box \langle \Diamond_i \rangle_{i<\alpha}) \in \Sigma$, of an *operation* over this carrier: $\mathbf{A}\,[\![O]\!]_{\mathrm{op}} : (\prod_{i<\alpha} \underline{\mathbf{A}}_{\Diamond_i}) \to \underline{\mathbf{A}}_{\Box}$.

*Example 8.* For any set $X$, define the $\mathtt{V}$-algebra $\mathbf{V}X$ by taking the carrier to be the set of countable (finite or infinite) $X$-subsets $\underline{\mathbf{V}X} \coloneqq \mathcal{P}^{\aleph_0}(X)$, and interpret choice as union $\mathbf{V}X[\![\bigvee_\alpha]\!]_{\mathrm{op}}\langle D_i \rangle_{i<\alpha} \coloneqq \bigcup_{i<\alpha} D_i$. □

*Example 9.* Define the $\mathtt{M}$-algebra $\mathbf{M}$ by taking the carrier to be the set of real-valued matrices of the corresponding dimensions, $\underline{\mathbf{M}}_{\mathbf{Hom}(m,n)} \coloneqq \mathbb{M}^{\mathbb{R}}_{m \times n}$, interpret the identity $\mathbf{M}[\![\mathrm{Id}_n]\!]_{\mathrm{op}} \coloneqq I_n \in \mathbb{M}^{\mathbb{R}}_{n \times n}$ as the identity matrix, and composition $\mathbf{M}[\![(\circ)]\!]_{\mathrm{op}} \coloneqq (\cdot)$ as matrix multiplication.

Let $\mathbf{A}$ be an $\mathtt{M}$-algebra. Define the *opposite* algebra $\mathbf{A}^{\mathsf{op}}$ by exchanging dimensions. So $\underline{\mathbf{A}^{\mathsf{op}}_{\mathbf{Hom}(m,n)}} \coloneqq \underline{\mathbf{A}}_{\mathbf{Hom}(n,m)}$, the same identity $\mathbf{A}^{\mathsf{op}}[\![\mathrm{Id}_n]\!]_{\mathrm{op}} \coloneqq \mathbf{A}[\![\mathrm{Id}_n]\!]_{\mathrm{op}}$, and reversing composition $\mathbf{A}^{\mathsf{op}}[\![(\circ)]\!]_{\mathrm{op}}(A, B) \coloneqq \mathbf{A}[\![(\circ)]\!]_{\mathrm{op}}(B, A)$. □

*Example 10 (term algebra).* The $\Sigma$-terms with variables from $\boldsymbol{X}$ carry a canonical algebra structure $\mathbf{F}^\Sigma \boldsymbol{X}$, given by $\underline{\mathbf{F}^\Sigma \boldsymbol{X}} \coloneqq \mathrm{Term}^\Sigma \boldsymbol{X}$, with each $O$-term constructor as the corresponding $O$-operation: $(\mathbf{F}^\Sigma \boldsymbol{X})\,[\![O]\!]_{\mathrm{op}}\,\langle t_i \rangle_i \coloneqq O\,\langle t_i \rangle_i$. □

A $\Sigma$-*algebra homomorphism* $\varphi : \mathbf{A} \to \mathbf{B}$ is a sorted-function $\varphi : \underline{\mathbf{A}} \to \underline{\mathbf{B}}$ that preserves the operations: $\varphi_{\Box}(\mathbf{A}\,[\![O]\!]_{\mathrm{op}}\,(a_1, \ldots, a_\alpha)) = \mathbf{B}\,[\![O]\!]_{\mathrm{op}}\,(\varphi_{\Diamond_1} a_1, \ldots, \varphi_{\Diamond_\alpha} a_\alpha)$.

*Example 11.* Transposing real-valued matrices $(-)^{\top} : \mathbb{M}^{\mathbb{R}}_{m\times n} \to \mathbb{M}^{\mathbb{R}}_{n\times m}$ is a homomorphism $(-)^{\top} : \mathbf{M} \to \mathbf{M}^{\mathsf{op}}$, by the well-known identity $(A \cdot B)^{\top} = B^{\top} \cdot A^{\top}$. $\square$

A $\Sigma$-algebra allows us to interpret every $\Sigma$-term, by assigning values to its variables. Formally, let $\mathbf{A}$ be a $\Sigma$-algebra. An $\boldsymbol{X}$-*environment in* $\mathbf{A}$ is a sorted function $e : \boldsymbol{X} \to \underline{\mathbf{A}}$. Given such an environment, interpret terms by induction:

$$\mathbf{A}\,[\![\boldsymbol{X} \vdash_{\Sigma} x : \boxbox]\!]_{\text{term}}\, e \coloneqq e_{\boxbox} x \qquad \mathbf{A}\,\left[\!\!\left[O\,\langle t_i\rangle_i\right]\!\!\right]_{\text{term}}\, e \coloneqq \mathbf{A}\,[\![O]\!]_{\text{op}}\,\left\langle \mathbf{A}\,[\![t_i]\!]_{\text{term}}\, e\right\rangle_i$$

*Example 12 (substitution).* An $\boldsymbol{X}$-environment in $\mathbf{F}^{\Sigma}\boldsymbol{X}$ amounts to a substitution, and interpreting terms in $\mathbf{F}^{\Sigma}\boldsymbol{X}$ amounts to substitution. $\square$

*Example 13 (evaluation homomorphism).* Evaluation using an $\boldsymbol{X}$-environment $e : \boldsymbol{X} \to \underline{\mathbf{A}}$ in a $\Sigma$-algebra $\mathbf{A}$ is a homomorphism $\mathbf{A}[\![-]\!]_{\text{term}} e : \mathbf{F}^{\Sigma}\boldsymbol{X} \to \mathbf{A}$. $\square$

A $\Sigma$-algebra $\mathbf{A}$ *validates* the equation $\boldsymbol{X} \vdash_{\Sigma} l = r : \boxbox$ when evaluation in all environments equates its sides: $\mathbf{A}[\![l]\!]_{\text{term}} e = \mathbf{A}[\![r]\!]_{\text{term}} e$ for all $e : \boldsymbol{X} \to \underline{\mathbf{A}}$. We then write $\mathbf{A} \vdash \boldsymbol{X} \vdash_{\Sigma} l = r : \boxbox$. A $\mathfrak{p}$-*model* is an algebra validating all of $\mathrm{Ax}_{\mathfrak{p}}$. The soundness theorem of equational logic states that every $\mathfrak{p}$-model validates all the equations in the algebraic theory of $\mathfrak{p}$.

*Example 14.* Referring to previous examples, the algebras $\mathbf{V}X$ are $\mathsf{V}$-models, the algebras $\mathbf{M}$ and $\mathbf{M}^{\mathsf{op}}$ are $\mathsf{M}$-models, and algebras of terms are $\emptyset$-models. $\square$

*Example 15.* Consider the $\Sigma_{\mathtt{J}}$-algebra $\mathbf{A}$ for which the carrier is the set of natural numbers $\underline{\mathbf{A}} \coloneqq \mathbb{N}$, join interprets as addition $\mathbf{A}[\![\vee]\!]_{\text{op}}(m, n) \coloneqq m + n$, and bottom as zero $\mathbf{A}[\![\bot]\!]_{\text{op}} \coloneqq 0$. This is *not* a $\mathsf{J}$-model, since, taking $e : \{x : \star\} \to \underline{\mathbf{A}}$ with $ex = 1$, we get $\mathbf{A}[\![x \vee x]\!]_{\text{term}} e \neq \mathbf{A}[\![x]\!]_{\text{term}} e$; and so $\mathbf{A} \not\vdash x : \star \vdash_{\mathtt{J}} x \vee x = x : \star$. $\square$

We end this section with representations of free models. These are $\mathfrak{p}$-models whose elements represent the $\Sigma_{\mathfrak{p}}$-terms up-to provable equality in $\mathfrak{p}$.

A $\mathfrak{p}$-*model* $\langle \mathbf{A}, e\rangle$ *over a family* $\boldsymbol{X}$ consists of a $\mathfrak{p}$-model $\mathbf{A}$ and an $\boldsymbol{X}$-environment in it $e : \boldsymbol{X} \to \underline{\mathbf{A}}$. A *free* $\mathfrak{p}$-model $\langle \mathbf{A}, \text{return}\rangle$ over a family $\boldsymbol{X}$ is then a $\mathfrak{p}$-model over $\boldsymbol{X}$ such that every environment in every $\mathfrak{p}$-model $e : \boldsymbol{X} \to \underline{\mathbf{B}}$ extends uniquely along return to a $\mathfrak{p}$-homomorphism $e^{\#} : \mathbf{A} \to \mathbf{B}$, i.e., for all $x \in \boldsymbol{X}_{\boxbox}$, we have: $e^{\#}_{\boxbox}(\text{return}_{\boxbox}\, a) = ea$. We then say that the algebra $\mathbf{A}$ *represents* $\boldsymbol{X}$-environments via the assignment $e \mapsto e^{\#}$, the corresponding *representation.*

The algebraic theory of effects [35] emphasises the role free models play in denotational semantics for programming languages with effects. In particular, given a free $\mathfrak{p}$-model over $\boldsymbol{X}$ for every family $\boldsymbol{X}$, one standardly obtains a monad suitable for the denotational semantics of a language with computational effects conforming to the operators in $\mathfrak{p}$.

*Example 16.* For any set $X$, the $\mathsf{V}$-algebra $\mathbf{V}X$ given by the countable powerset in example 8 represents $X$-environments; together with return $x \coloneqq \{x\}$ it forms a free $\mathsf{V}$-model over $X$. The representation assigns $e : X \to \underline{\mathbf{B}}$ to $e^{\#} : \mathbf{V}X \to \mathbf{B}$, defined $e^{\#}D \coloneqq \mathbf{B}[\![\bigvee_{|D|}]\!]_{\text{op}}\langle ex\rangle_{x\in D}$; how it enumerates $D$ doesn't matter since $\mathbf{B}$ is a $\mathsf{V}$-model. The data $\langle X \mapsto \underline{\mathbf{V}X}, \text{return}, (-)^{\#}\rangle$ is a monad. $\square$

## 4 Shared state

To define the equational theory of shared state, we first recall the standard, single sorted *(non-deterministic) global state* theory $\mathsf{G}$ [17, 29, 35]. The variant we present here has countable non-determinism, and the global state operators manipulate a common memory store $\mathbb{S} \coloneqq \mathbb{L} \to \mathbb{B}$ with a finite set of locations $\mathbb{L} \neq \emptyset$ each storing a bit $\mathbb{B} \coloneqq \{\mathtt{0}, \mathtt{1}\}$. A larger finite set of storable-values would not be conceptually different. Infinite sets of storable-values or locations work similarly with more involved representation theorems. In concrete examples, we let $\mathbb{L} = \{\mathtt{x}, \mathtt{y}\}$ and use non-bracketed vectors for stores, e.g. $\substack{\mathtt{1}\\ \mathtt{0}}$ denotes $\left(\substack{\mathtt{x}\mapsto\mathtt{1}\\ \mathtt{y}\mapsto\mathtt{0}}\right)$.

The signature $\Sigma_{\mathsf{G}}$ consists of the countable-join semilattice operators (example 2), as well as two kinds of memory-access operators: *lookup* operators $\mathsf{L}_{\ell} : 2$, to look a location $\ell \in \mathbb{L}$ up and branch according to the value found; and *update* operators $\mathsf{U}_{\ell,b} : 1$, to update a location $\ell \in \mathbb{L}$ to the value $b \in \mathbb{B}$. The global state axioms $\mathrm{Ax}_{\mathsf{G}}$ consists of the countable-join semilattice axioms (example 5), as well as the following:

**Non-deterministic global state (omitting semilattice axioms)**

| | | | |
|---|---|---|---|
| ($\mathtt{UL}$) | $\mathsf{U}_{\ell,b}\,\mathsf{L}_{\ell}(x_{\mathtt{0}}, x_{\mathtt{1}}) = \mathsf{U}_{\ell,b}\,x_b$ | ($\mathtt{LU}$) | $\mathsf{L}_{\ell}(\mathsf{U}_{\ell,\mathtt{0}}\,x, \mathsf{U}_{\ell,\mathtt{1}}\,x) = x$ |
| ($\mathtt{UU}$) | $\mathsf{U}_{\ell,b'}\,\mathsf{U}_{\ell,b}\,x = \mathsf{U}_{\ell,b}\,x$ | (ND-$\mathtt{U}$) | $\bigvee_{i<\alpha} \mathsf{U}_{\ell,b}\,x_i = \mathsf{U}_{\ell,b} \bigvee_{i<\alpha} x_i$ |
| ($\mathtt{UUc}$) | $\mathsf{U}_{\ell,b}\,\mathsf{U}_{\ell',b'}\,x = \mathsf{U}_{\ell',b'}\,\mathsf{U}_{\ell,b}\,x$ | where $\ell \neq \ell'$ | |

The induced algebraic theory $\mathsf{G}$ includes axioms of less succinct presentations of the same theory [29]. For example, lookup also distributes over binary join, so the theory admits inequational reasoning; consecutively looking the same location up can be merged, e.g. $x_{\mathtt{0}}, x_{\mathtt{1}}, y \vdash_{\mathsf{G}} \mathsf{L}_{\ell}(\mathsf{L}_{\ell}(x_{\mathtt{0}}, x_{\mathtt{1}}), y) = \mathsf{L}_{\ell}(x_{\mathtt{0}}, y)$; and other combinations of looking-up and updating different locations commute, e.g. for any $\ell \neq \ell'$ we have $x_{\mathtt{0}}, x_{\mathtt{1}} \vdash_{\mathsf{G}} \mathsf{L}_{\ell}(\mathsf{U}_{\ell',b}\,x_{\mathtt{0}}, \mathsf{U}_{\ell',b}\,x_{\mathtt{1}}) = \mathsf{U}_{\ell',b}\,\mathsf{L}_{\ell}(x_{\mathtt{0}}, x_{\mathtt{1}})$.

Our two-sorted presentation $\boldsymbol{\mathbb{S}}$ of *shared state* extends global state. Its sorts are $\mathbf{sort}_{\Sigma_{\boldsymbol{\mathbb{S}}}} = \{\bullet, \circ\}$. The *hold* sort ($\bullet$) represents an uninterrupted sequence of memory accesses, whereas the *cede* sort ($\circ$) allows control to pass to the environment. The operators and the arities of the signature $\Sigma_{\boldsymbol{\mathbb{S}}}$ consist of a copy of $\Sigma_{\mathsf{G}}$ at $\bullet$, a copy of $\Sigma_{\mathsf{V}}$ at $\circ$, and new operators $\lhd : \circ\langle\bullet\rangle$ and $\rhd : \bullet\langle\circ\rangle$.

The intuitive reading for algebraic effects is from the outside in. With this intuition, one interpretation of the operators $\lhd$ and $\rhd$ is to acquire and release a global lock. The hold sort ($\bullet$) represents the lock being held by one of the threads in the program. The cede sort ($\circ$) represents points in the execution in which one of the threads in the concurrent environment may acquire the lock. The sorts ensure exclusive access to the lock, and therefore to the store. In an alternative interpretation, these operators delimit atomic blocks; their sorts prevent nesting.

The shared state axioms $\mathrm{Ax}_{\boldsymbol{\mathbb{S}}}$ include a copy of the (non-deterministic) global state axioms $\mathrm{Ax}_{\mathsf{G}}$ at $\bullet$ and a copy of the countable-join semilattice axioms $\mathrm{Ax}_{\mathsf{V}}$ at $\circ$. In particular, $\boldsymbol{\mathbb{S}}$ proves the semi-lattice axioms in both sorts. It further includes standard strict distributivity axioms for the new unary operators:

**Strict distributivity of $\lhd$ and $\rhd$**

(ND-$\lhd$) $\bigvee_{i<\alpha} \lhd x_i = \lhd \bigvee_{i<\alpha} x_i$ (ND-$\rhd$) $\bigvee_{i<\alpha} \rhd x_i = \rhd \bigvee_{i<\alpha} x_i$

With these axioms, $\mathbb{S}$ supports inequational reasoning, which represents the semantic refinement relation used to validate program transformations [e.g. 12].

Finally, $\mathrm{Ax}_{\mathbb{S}}$ axiomatises $\lhd$ and $\rhd$ as an *(insertion)-closure pair* [e.g. 2]:

**Closure pair** (Empty) $\lhd \rhd y = y$ (Fuse) $\rhd \lhd x \geq x$

They are compatible with the global-lock interpretation:

**Empty ($\lhd \rhd y = y$).** Acquiring and immediately releasing the lock has no effect on the sequence of effects that can occur as a result of arbitrary interleavings.

**Fuse ($\rhd \lhd x \geq x$).** Releasing and immediately acquiring the lock only allows more behaviours. The environment may or may not interleave there.

To summerise, $\mathrm{Ax}_{\mathbb{S}} \coloneqq \mathrm{Ax}^{\bullet}_{\mathsf{G}} \cup \mathrm{Ax}^{\circ}_{\mathsf{V}} \cup \{\text{ND-}\rhd, \text{ND-}\lhd\} \cup \{\text{Empty}, \text{Fuse}\}$.

*Example 17.* The $\Sigma_{\mathbb{S}}$-equations appearing below are named after corresponding transformations that may or may not be valid, depending on the setting (e.g. is there concurrency, and under what assumptions), all $\circ$-sorted over $\{x : \circ\}$:

$$\lhd \mathsf{L}_{\ell}(\rhd x, \rhd x) = x \qquad \text{(Irrelevant Read Intro \& Elim)}$$

$$\lhd \mathsf{U}_{\ell,b_1} \rhd \lhd \mathsf{U}_{\ell,b_2} \rhd x \geq \lhd \mathsf{U}_{\ell,b_2} \rhd x \qquad \text{(Write Elim)}$$

$$\lhd \mathsf{U}_{\ell,b_1} \rhd \lhd \mathsf{U}_{\ell,b_2} \rhd x \leq \lhd \mathsf{U}_{\ell,b_2} \rhd x \qquad \text{(Write Intro)}$$

Intuitively, Irrelevant Read Intro & Elim should be valid in our setting, as looking a value up is not observable by the environment, and the computation itself disregards the value. Write Elim should be valid too, because it is possible that the environment does not look $\ell$ up at the interference point between the updates on the LHS, covering the behaviour denoted by the RHS. On the other hand, Write Intro should be invalid in our setting because only on the LHS can a concurrently running thread look $\ell$ up and find $b_1$. Formally, we will show $\mathbb{S}$ does not prove Write Intro in example 25. Here we show $\mathbb{S}$ proves the other two:

$$\lhd \mathsf{L}_{\ell}\left(\rhd x, \rhd x\right) \overset{\mathtt{LU}}{=} \lhd \mathsf{L}_{\ell}\left(\mathsf{U}_{\ell,0}\, \mathsf{L}_{\ell}\left(\rhd x, \rhd x\right), \mathsf{U}_{\ell,1}\, \mathsf{L}_{\ell}\left(\rhd x, \rhd x\right)\right)$$

$$\overset{\mathtt{UL}}{=} \lhd \mathsf{L}_{\ell}\left(\mathsf{U}_{\ell,0} \rhd x, \mathsf{U}_{\ell,1} \rhd x\right) \overset{\mathtt{LU}}{=} \lhd \rhd x \overset{\text{Empty}}{=} x$$

$$\lhd \mathsf{U}_{\ell,b_1} \rhd \lhd \mathsf{U}_{\ell,b_2} \rhd x \overset{\text{Fuse}}{\geq} \lhd \mathsf{U}_{\ell,b_1}\, \mathsf{U}_{\ell,b_2} \rhd x \overset{\mathtt{UU}}{=} \lhd \mathsf{U}_{\ell,b_2} \rhd x \qquad \square$$

## 5 Representation

We now establish the representation theorem describing a free $\mathbb{S}$-model over any $\boldsymbol{X} \in \mathbf{Set}^{\{\bullet,\circ\}}$. Following Brookes [7], we use sets of traces to denote behaviours.

### 5.1 Sorted traces

A *(sorted) trace* starts with a sort ($\bullet$ or $\circ$) followed by a non-empty sequence of state transitions, and ending in a sorted value. The initial sort in the trace and the initial store in each transition represent assumptions the trace relies on from its concurrent and sequential environment. The final sort and value and the final store in each transition represent guarantees the trace makes to its environment.

Formally, a *(state) transition* is a pair $\langle \sigma, \rho \rangle \in \mathbb{S} \times \mathbb{S}$. Let $\xi^{?} \in (\mathbb{S} \times \mathbb{S})^{*}$ range over possibly empty sequences of transitions, and $\xi \in (\mathbb{S} \times \mathbb{S})^{+}$ range over non-empty ones. For any set $X$, define the set of *$X$-valued Brookes traces* $\mathsf{T}X \coloneqq (\mathbb{S} \times \mathbb{S})^{+} \times X$, also used in Brookes's model (§6). For any family $\boldsymbol{X} \in \mathbf{Set}^{\{\bullet,\circ\}}$ define the $\{\bullet,\circ\}$-sorted family $\boldsymbol{\mathsf{T}}\boldsymbol{X}$ by $(\boldsymbol{\mathsf{T}}\boldsymbol{X})_{\boxdot} \coloneqq \mathsf{T} \oint \boldsymbol{X}$; and define the set of *(sorted) traces over $\boldsymbol{X}$* by:

$$\mathbb{T}\boldsymbol{X} \coloneqq \oint \boldsymbol{\mathsf{T}}\boldsymbol{X} = \{\bullet,\circ\} \times (\mathbb{S} \times \mathbb{S})^{+} \times \coprod_{\Diamonddot \in \{\bullet,\circ\}} \boldsymbol{X}_{\Diamonddot}$$

A *$\boxdot$-sorted $\Diamonddot$-valued trace* is one of the form $\boxdot\xi\Diamonddot x \coloneqq \langle \boxdot, \xi, \mathrm{in}_{\Diamonddot}\, x \rangle$ in the set $\mathbb{T}\boldsymbol{X}$.

*Example 18.* $\bullet\langle {}^{1}_{1}, {}^{1}_{0} \rangle \langle {}^{1}_{1}, {}^{0}_{0} \rangle \circ 7 \in \mathbb{T}\boldsymbol{X}$, with $\boldsymbol{X}_{\circ} = \mathbb{N}$, is $\bullet$-sorted and $\circ$-valued. □

Intuitively, the trace $\boxdot\xi\Diamonddot x$ models a potential behaviour, or protocol, that a shared-state program phrase under preemptive interleaving concurrency can exhibit, or adhere to, given as a rely/guarantee sequence.

*Example 19.* The behaviour denoted by $\bullet\langle {}^{1}_{1}, {}^{1}_{0} \rangle \langle {}^{1}_{1}, {}^{0}_{0} \rangle \circ 7$ relies on the preceding environment for ${}^{1}_{1}$ and for the sequential environment to hold access to the store; then guarantees ${}^{1}_{0}$; then relies on ${}^{1}_{1}$; and finally guarantees ${}^{0}_{0}$, and returns 7 to the succeeding sequential environment, ceding exclusive store access. □

One can make these trace-semantic concepts more formal, for example, when formulating an adequacy proof w.r.t. an operational semantics. We will not define these concepts formally since we will not need the additional level of rigour, for example, because we appeal to the well-established adequacy of Brookes's model.

We implicitly understand the exclusive access to the store is ceded ($\circ$) between transitions. For example, for the trace $\bullet\langle {}^{1}_{1}, {}^{1}_{0} \rangle \langle {}^{1}_{1}, {}^{0}_{0} \rangle \circ 7$, we could write $\bullet\langle {}^{1}_{1}, {}^{1}_{0} \rangle \circ \langle {}^{1}_{1}, {}^{0}_{0} \rangle \circ 7$ for emphasis. A hypothetical $\bullet\langle {}^{1}_{1}, {}^{1}_{0} \rangle \bullet \langle {}^{1}_{1}, {}^{0}_{0} \rangle \circ 7$ would denote an impossible behaviour, making intermediate sorts redundant.

One of Brookes's innovations is that sets of traces should be closed under what we now call *(trace) deductions*. Specifically, Brookes identified two such deductions, given as binary relations called $\mathsf{stutter}$ ($\xrightarrow{\mathsf{st}}$) and $\mathsf{mumble}$ ($\xrightarrow{\mathsf{mu}}$), defined in such a way that if the program phrase can adhere to the source protocol (left of arrow), then it can adhere to the target protocol (right of arrow).

We define these deductions in our two-sorted setting. For convenience, we write $\boxdot\xi_1^{?}\circ\xi_2^{?}\Diamonddot x$ for the trace $\boxdot\xi_1^{?}\xi_2^{?}\Diamonddot x$ in which, intuitively, the lock is ceded ($\circ$) at the marked spot. Formally, we require that both (a) if $\xi_1^{?}$ is empty, then $\boxdot = \circ$; and (b) if $\xi_2^{?}$ is empty, then $\Diamonddot = \circ$. In particular, the requirement holds when both $\xi_1^{?}$ and $\xi_2^{?}$ are non-empty, where we implicitly assume the ceded sort between them; and in the case of a $\circ$-sorted $\circ$-valued trace, i.e. $\boxdot = \circ = \Diamonddot$.

*Example 20.* We have the following valid/invalid notations for $\bullet\langle \substack{1\\1}, \substack{1\\0}\rangle\langle \substack{1\\1}, \substack{0\\0}\rangle\circ 7$:

$$\text{valid: } \bullet\langle \substack{1\\1}, \substack{1\\0}\rangle\circ\langle \substack{1\\1}, \substack{0\\0}\rangle\circ 7 \quad \bullet\langle \substack{1\\1}, \substack{1\\0}\rangle\langle \substack{1\\1}, \substack{0\\0}\rangle\circ\circ 7 \qquad \text{invalid: } \bullet\circ\langle \substack{1\\1}, \substack{1\\0}\rangle\langle \substack{1\\1}, \substack{0\\0}\rangle\circ 7 \qquad \square$$

We define the following *sorted* stutter *and* mumble *deductions*:

$$\Box\xi_1^?\circ\xi_2^?\Diamond x \xrightarrow{\mathtt{st}} \Box\xi_1^?\langle\sigma,\sigma\rangle\xi_2^?\Diamond x \qquad \Box\xi_1^?\langle\sigma,\rho\rangle\langle\rho,\theta\rangle\xi_2^?\Diamond x \xrightarrow{\mathtt{mu}} \Box\xi_1^?\langle\sigma,\theta\rangle\xi_2^?\Diamond x$$

The condition on stutter's source rules out deductions which implicitly cede access to the store to the concurrent environment at the ends of the trace. We will compare these deductions to Brookes's in §6.

*Example 21.* These deductions are valid, highlighting the change to the trace:

$$\bullet\langle \substack{1\\1}, \substack{1\\0}\rangle\langle \substack{1\\1}, \substack{0\\0}\rangle\circ 7 \xrightarrow{\mathtt{st}} \bullet\langle \substack{1\\1}, \substack{1\\0}\rangle\langle \substack{1\\1}, \substack{0\\0}\rangle\langle \substack{0\\1}, \substack{0\\1}\rangle\circ 7 \qquad \bullet\langle \substack{1\\1}, \substack{1\\0}\rangle\langle \substack{1\\0}, \substack{0\\0}\rangle\circ 7 \xrightarrow{\mathtt{mu}} \bullet\langle \substack{1\\1}, \substack{0\\0}\rangle\circ 7$$

However, thanks to the condition on stutter's source, this deduction is invalid:

$$\bullet\langle \substack{1\\1}, \substack{1\\0}\rangle\langle \substack{1\\1}, \substack{0\\0}\rangle\circ 7 \not\xrightarrow{\mathtt{st}} \bullet\langle \substack{0\\1}, \substack{0\\1}\rangle\langle \substack{1\\1}, \substack{1\\0}\rangle\langle \substack{1\\1}, \substack{0\\0}\rangle\circ 7$$

The source protocol relies on the preceding sequential environment for $\substack{1\\1}$. We prohibit relaxing the protocol to rely on the concurrent environment for it. $\square$

The stutter and mumble deductions follow the rely/guarantee intuition:

**Stuttering** ($\Box\xi_1^?\circ\xi_2^?\Diamond x \xrightarrow{\mathtt{st}} \Box\xi_1^?\langle\sigma,\sigma\rangle\xi_2^?\Diamond x$) means a thread-pool also obeys the protocol that guarantees a state $\sigma$ by relying on its environment for $\sigma$.

**Mumbling** ($\Box\xi_1^?\langle\sigma,\rho\rangle\langle\rho,\theta\rangle\xi_2^?\Diamond x \xrightarrow{\mathtt{mu}} \Box\xi_1^?\langle\sigma,\theta\rangle\xi_2^?\Diamond x$) means a thread-pool that guarantees the store $\rho$ it later relies on also obeys the protocol in which we exclude the environment's access to the store $\rho$ at that point.

Sets of traces represent a non-deterministic choice between the behaviours that a program phrase may exhibit. For such a set $K$, define its *closure* under trace deduction $K^\dagger$ as the least set $K'$ such that: $K \subseteq K'$; and if $\tau_1 \in K'$ and $\tau_1 \xrightarrow{\mathtt{x}} \tau_2$ for $\mathtt{x} \in \{\mathtt{st}, \mathtt{mu}\}$, then $\tau_2 \in K'$. According to the rely/guarantee intuition above, a program phrase that is compatible with a set of traces is also compatible with its closure. We therefore represent program phrases as *closed* sets, i.e. sets $K$ such that $K = K^\dagger$. The closure $K^\dagger$ of a countable $K$ is countably infinite—by stuttering indefinitely—unless $K$ is a finite set of single-transition $\bullet$-sorted $\bullet$-valued traces, in which case $K$ is already closed.

For a set of traces $U$ and sort $\Box \in \{\bullet, \circ\}$, define a $\{\bullet, \circ\}$-sorted family $\mathcal{P}^{\aleph_0}(U)$ by taking its $\Box$ component to be the set $\mathcal{P}^{\aleph_0}_{\Box}(U)$ of countable subsets of $U$ whose elements are all $\Box$-sorted. Similarly, define $\mathcal{P}^{\dagger}_{\Box}(U) \subseteq \mathcal{P}^{\aleph_0}_{\Box}(U)$ to be the set of *closed* countable subsets of $U$ whose elements are all $\Box$-sorted.

The *prefixing* function adds the given transition to each $\bullet$-sorted trace:

$$(\sigma,\rho) : \mathcal{P}^{\aleph_0}_{\bullet}(\mathbb{T}\boldsymbol{X}) \to \mathcal{P}^{\aleph_0}_{\bullet}(\mathbb{T}\boldsymbol{X}) \qquad (\sigma,\rho)\,K \coloneqq \left\{\bullet\langle\sigma,\theta\rangle\xi^?\Diamond x \;\middle|\; \bullet\langle\rho,\theta\rangle\xi^?\Diamond x \in K\right\}$$

It lifts to closed sets, i.e. $K \in \mathcal{P}^{\dagger}_{\bullet}(\mathbb{T}\boldsymbol{X})$ implies that $(\sigma,\rho)\,K \in \mathcal{P}^{\dagger}_{\bullet}(\mathbb{T}\boldsymbol{X})$.

### 5.2 Representation theorem

For $\boldsymbol{X} \in \mathbf{Set}^{\{\bullet,\circ\}}$, define the $\Sigma_{\$}$-algebra of *$\boldsymbol{X}$-valued closed trace-sets* $\mathbf{R}\boldsymbol{X}$ as:

$$\underline{\mathbf{R}\boldsymbol{X}}_{\square} \coloneqq \mathcal{P}^{\dagger}_{\square}(\mathbb{T}\boldsymbol{X}) \qquad [\![\mathsf{U}_{\ell,b}]\!]_{\mathrm{op}} K \coloneqq \bigcup\nolimits_{\sigma\in\mathbb{S}} (\sigma, \sigma[\ell \mapsto b])\, K$$

$$[\![\textstyle\bigvee_{i<\alpha}]\!]_{\mathrm{op}} K_i \coloneqq \bigcup\nolimits_{i<\alpha} K_i \qquad [\![\mathsf{L}_{\ell}]\!]_{\mathrm{op}}(K_0, K_1) \coloneqq \bigcup\nolimits_{\sigma\in\mathbb{S}} (\sigma,\sigma)\, K_{\sigma_\ell}$$

$$[\![\triangleleft]\!]_{\mathrm{op}} K \coloneqq \{\circ\xi\Diamond x \mid \bullet\xi\Diamond x \in K\}^{\dagger} \qquad [\![\triangleright]\!]_{\mathrm{op}} K \coloneqq \left\{\bullet\langle\sigma,\sigma\rangle\xi\Diamond x \;\middle|\; \sigma\in\mathbb{S}, \circ\xi\Diamond x \in K\right\}^{\dagger}$$

Additionally, define $\mathrm{return} : \boldsymbol{X} \to \underline{\mathbf{R}\boldsymbol{X}}$ by $\mathrm{return}_{\square}\, x \coloneqq \{\square\langle\sigma,\sigma\rangle\square x \mid \sigma \in \mathbb{S}\}^{\dagger}$.

The rest of this section establishes that the algebra $\langle \mathbf{R}\boldsymbol{X}, \mathrm{return}\rangle$ over $\boldsymbol{X}$ is a free $\$$-model over $\boldsymbol{X}$. A key ingredient is *reification*: for any $\{\bullet,\circ\}$-sorted family $\boldsymbol{X}$, we define a sorted-function $\mathrm{reify} : \mathcal{P}^{\aleph_0}(\mathbb{T}\boldsymbol{X}) \to \mathrm{Term}^{\Sigma_{\$}}\boldsymbol{X}$, choosing a representative term $t_2 \coloneqq \mathrm{reify}[\![\boldsymbol{X} \vdash t_1]\!]_{\mathrm{term}}$ such that $\boldsymbol{X} \vdash_{\$} t_1 = t_2$. This use of countable choice is inessential, the mere existence of the defining term $t_2$ suffices.

First define for any $\ell \in \mathbb{L}$ and $b \in \mathbb{B}$ the *cell assertion* term $x : \bullet \vdash_{\Sigma_{\$}} \mathsf{A}_{\ell,b}\, x : \bullet$ that looks $\ell$ up and only continues if it holds $b$:

$$x : \bullet \vdash_{\Sigma_{\$}} \mathsf{A}_{\ell,0}\, x \coloneqq \mathsf{L}_{\ell}(x, \bot) : \bullet \qquad x : \bullet \vdash_{\Sigma_{\$}} \mathsf{A}_{\ell,1}\, x \coloneqq \mathsf{L}_{\ell}(\bot, x) : \bullet$$

Next, for any $\sigma, \rho \in \mathbb{S}$ we define the *open transition* $x : \bullet \vdash_{\Sigma_{\$}} \wr\sigma,\rho\wr\, x : \bullet$, as a term that asserts the state is $\sigma$, then updates the state to $\rho$, and returns $x$:

$$x : \bullet \vdash_{\Sigma_{\$}} \wr\sigma,\rho\wr\, x \coloneqq \mathsf{A}_{\mathtt{l}_1,\sigma_{\mathtt{l}_1}} \ldots \mathsf{A}_{\mathtt{l}_n,\sigma_{\mathtt{l}_n}}\, \mathsf{U}_{\mathtt{l}_1,\rho_{\mathtt{l}_1}} \ldots \mathsf{U}_{\mathtt{l}_n,\rho_n}\, x : \bullet \qquad (\mathbb{L} = \{\mathtt{l}_1, \ldots, \mathtt{l}_n\})$$

Now we can represent traces as terms. Define the $\Sigma_{\$}$-term reifying a trace $x : \Diamond \vdash_{\Sigma_{\$}} \underline{\square\xi\Diamond x} : \square$ by sequencing open transition as they are in $\xi$, separated by $\triangleright\triangleleft$; and delimited by $\triangleleft$ on the left if $\square = \circ$ and by $\triangleright$ on the right if $\Diamond = \circ$.

*Example 22.* $x : \circ \vdash_{\Sigma_{\$}} \underline{\bullet\langle\sigma,\rho\rangle\langle\sigma',\rho'\rangle\circ x} \coloneqq \wr\sigma,\rho\wr \triangleright\triangleleft \wr\sigma',\rho'\wr \triangleright x : \bullet$ □

Trace deductions are sound w.r.t. this encoding, in the following sense:

**Proposition 23.** *Assume that $\tau_1$ and $\tau_2$ are $\square$-sorted traces over $\{x : \Diamond\}$, such that $\tau_1 \xrightarrow{\mathtt{x}} \tau_2$ for $\mathtt{x} \in \{\mathtt{st}, \mathtt{mu}\}$. Then $x : \Diamond \vdash_{\Sigma_{\$}} \underline{\tau_1} \geq \underline{\tau_2} : \square$.*

Finally, we reify a trace set by reifying its traces in a chosen enumeration:

$$\mathrm{reify} : \mathcal{P}^{\aleph_0}(\mathbb{T}\boldsymbol{X}) \to \mathrm{Term}^{\Sigma_{\$}}\boldsymbol{X} \qquad \mathrm{reify}_{\square}\, K \coloneqq \left(\boldsymbol{X} \vdash_{\Sigma_{\$}} \textstyle\bigvee_{\tau\in K} \underline{\tau} : \square\right)$$

By proposition 23, closure preserves reification: $\boldsymbol{X} \vdash_{\$} \mathrm{reify}_{\square}\, K = \mathrm{reify}_{\square}\, K^{\dagger} : \square$.

Using reification, we state the representation theorem (proof in §C).

**Theorem 24 ($\$$-representation).** *The pair $\langle \mathbf{R}\boldsymbol{X}, \mathrm{return}\rangle$ is a free $\$$-model over $\boldsymbol{X}$. Its representation sends environments $e : \boldsymbol{X} \to \underline{\mathbf{A}}$ to $\$$-homomorphisms $e^{\#} : \mathbf{R}\boldsymbol{X} \to \mathbf{A}$ by $e^{\#}_{\square} K \coloneqq \mathbf{A}[\![\mathrm{reify}_{\square}\, K]\!]_{\mathrm{term}} e$. Moreover, for $\mathbf{A} = \mathbf{R}\boldsymbol{Y}$ we have:*

$$e^{\#}_{\square} K = \left\{ \square\xi_1\xi_2\Diamond y \;\middle|\; \begin{matrix} \square\xi_1\circ x \in K, \\ \circ\xi_2\Diamond y \in e_{\circ} x \end{matrix} \right\}^{\dagger} \cup \left\{ \square\xi^{?}_1\langle\sigma,\theta\rangle\xi^{?}_2\Diamond y \;\middle|\; \begin{matrix} \square\xi^{?}_1\langle\sigma,\rho\rangle\bullet x \in K, \\ \bullet\langle\rho,\theta\rangle\xi^{?}_2\Diamond y \in e_{\bullet} x \end{matrix} \right\}^{\dagger}.$$

*Example 25.* The model $\mathbf{R}\{x : \circ\}$ invalidates Write Intro:

$$\mathbf{R}\{x : \circ\}[\![\triangleleft\, \mathsf{U}_{\ell,b_1} \triangleright\triangleleft\, \mathsf{U}_{\ell,b_2} \triangleright x]\!]_{\mathrm{term}}\mathrm{return} \neq \mathbf{R}\{x : \circ\}[\![\triangleleft\, \mathsf{U}_{\ell,b_2} \triangleright x]\!]_{\mathrm{term}}\mathrm{return}$$

Every trace in the right-hand set has at most one state-changing transition. The left-hand set has traces with two. Therefore, $\$$ does not prove Write Intro. □

# 6 Recovering Brookes's model

The theory $\mathbb{S}$ recovers Brookes's model (§6.1). We recover it twice, using different strategies that offer different perspectives. The first transforms the monad induced by the representation of §5.2 along a right adjoint $(-)_{\circ} : \mathbf{Set}^{\{\bullet,\circ\}} \to \mathbf{Set}$ sending each $\{\bullet,\circ\}$-family $\boldsymbol{X}$ to the set $\boldsymbol{X}_{\circ} \coloneqq \{x \mid (x : \circ) \in \boldsymbol{X}\}$ (§6.2). In the second, we define a single-sorted theory of transitions $\mathsf{B}$ that recovers Brookes's model straightforwardly (§6.3). In this theory, the transition operators correspond to Brookes's `await` construct. After swiftly introducing translations (§6.4), we show that $\mathsf{B}$ translation embeds into $\mathbb{S}$. The translation factors through another, two-sorted, theory of transitions $\mathsf{Tr}$ (§6.5).

## 6.1 Brookes's model

We designed our notions of traces, deduction, etc. from §5.1 based on the following model of Brookes [7], in which traces cannot hold exclusive memory access at their ends. In this model, ceding access is implicit.

For any set $X \in \mathbf{Set}$, recall the set of Brookes traces $\mathsf{T}X \coloneqq (\mathbb{S} \times \mathbb{S})^{+} \times X$ from §5.1. Writing $\xi x$ for $\langle \xi, x \rangle$, Brookes's stutter and mumble deductions are:

$$\xi_1^{?}\xi_2^{?}x \xrightarrow{\texttt{st}} \xi_1^{?}\langle\sigma,\sigma\rangle\xi_2^{?}x \qquad \xi_1^{?}\langle\sigma,\rho\rangle\langle\rho,\theta\rangle\xi_2^{?}x \xrightarrow{\texttt{mu}} \xi_1^{?}\langle\sigma,\theta\rangle\xi_2^{?}x$$

We reuse the notation $(-)^{\dagger}$ for closure under these deductions.

The difference between Brookes's deductions and our multi-sorted deductions is the maintenance of the sort on each end of the trace. In particular, Brookes's stutter does not need to explicitly allow interleaving at the relevant position in the source, because the environment may always interleave on either end.

Brookes's semantic domain $BX \coloneqq \mathcal{P}^{\dagger}(\mathsf{T}X)$ forms a monad. The monadic unit is $\mathrm{return} : X \to BX$, $\mathrm{return}\, x \coloneqq \{\langle\sigma,\sigma\rangle x \mid \sigma \in \mathbb{S}\}^{\dagger}$. The Kleisli extension $e^{\#} : BX \to BY$ of every $e : X \to BY$ is $e^{\#}K \coloneqq \{\xi_1\xi_2 y \mid \xi_1 x \in K, \xi_2 y \in ex\}^{\dagger}$. It interprets memory accesses, dereferencing $(\ell!)$ and mutation $(\ell \coloneqq b)$, as follows:

$$[\![\ell!]\!] : \mathbb{1} \xrightarrow{\{\langle\sigma,\sigma\rangle\sigma_\ell \mid \sigma\in\mathbb{S}\}^{\dagger}} B\mathbb{B} \qquad [\![\ell \coloneqq b]\!] : \mathbb{1} \xrightarrow{\{\langle\sigma,\sigma[\ell\mapsto b]\rangle\langle\rangle \mid \sigma\in\mathbb{S}\}^{\dagger}} B\mathbb{1}$$

These *generic effects* [36] correspond to these monadic algebraic operations:

$$\begin{aligned} [\![\mathsf{R}_\ell]\!] &: (BX)^2 \to BX & \quad [\![\mathsf{R}_\ell]\!](K_0, K_1) &\coloneqq \{\langle\sigma,\sigma\rangle\xi x \mid \sigma \in \mathbb{S}, \xi x \in K_{\sigma_\ell}\}^{\dagger} \\ [\![\mathsf{W}_{\ell,b}]\!] &: BX \to BX & \quad [\![\mathsf{W}_{\ell,b}]\!]K &\coloneqq \{\langle\sigma,\sigma[\ell \mapsto b]\rangle\xi x \mid \sigma \in \mathbb{S}, \xi x \in K\}^{\dagger} \end{aligned}$$

## 6.2 Recovery via an adjunction

In Brookes's model, yielding to the concurrent environment is implicit, and always allowed. From our two-sorted point-of-view, we expect the traces in Brookes's model to represent $\circ$-sorted $\circ$-valued traces.

There is an abstract construction that recovers the monad and its operations in §6.2 from our $\{\bullet,\circ\}$-sorted model. The functor $(-)_{\circ} : \mathbf{Set}^{\{\bullet,\circ\}} \to \mathbf{Set}$

has a left-adjoint $(-)^{\circ} : \mathbf{Set} \to \mathbf{Set}^{\{\bullet,\circ\}}$. This functor sends each set $X$ to the $\{\bullet, \circ\}$-family $X^{\circ} \coloneqq \{x : \circ \mid x \in X\}$, using the set-like notation for families we introduced in §3.1. Monads transform along adjoints, and transforming the monad obtained standardly from the representation of §5.2 along the adjunction above results in Brookes's model. Explicitly, denoting $B_{\circ}X \coloneqq \underline{\mathbf{R}X^{\circ}}_{\circ} = \mathcal{P}^{\dagger}_{\circ}(\mathbb{T}X^{\circ})$, the resulting monad over $\mathbf{Set}$ is $\langle B_{\circ}, \mathrm{return}_{\circ}, (-)^{\#}_{\circ}\rangle$. This monad is isomorphic to Brookes's $\langle B, \mathrm{return}, (-)^{\#}\rangle$ above by way of removing $\circ$ from both ends of every trace. Thus, the Brookes model amounts to the free $\boldsymbol{\mathbb{S}}$-model from §5.2 transformed along the adjunction $(-)^{\circ} \dashv (-)_{\circ}$. The monad $\mathbf{R}$ supports the following generic effects. The adjunction transforms them, via its natural bijection on homsets, into Brookes's generic effects for memory access:

$$[\![\ell!]\!] : \mathbb{1}^{\circ} \xrightarrow{[\![\lhd\, \mathsf{L}_{\ell}(\rhd\, 0, \rhd\, 1)]\!]} \mathbf{R}\mathbb{B}^{\circ} \qquad [\![\ell \coloneqq b]\!] : \mathbb{1}^{\circ} \xrightarrow{[\![\lhd\, \mathsf{U}_{\ell,b} \rhd \langle\rangle]\!]} \mathbf{R}\mathbb{1}^{\circ}$$

### 6.3 The single-sorted theory of transitions

There is a more direct, single-sorted presentation $\mathsf{B}$ for Brookes's model. It uses transitions as operators rather than lookup and update operators. The signature $\Sigma_{\mathsf{B}}$ consists of countable-joins $\Sigma_{\mathsf{V}}$ and a unary transition operator $\langle\sigma, \rho\rangle$ for every $\sigma, \rho \in \mathbb{S}$. The axioms $\mathrm{Ax}_{\mathsf{B}}$ consist of the countable-join semilattice axioms $\mathrm{Ax}_{\mathsf{V}}$, strict distributivity axioms (ND-$\mathsf{B}$) $\langle\sigma, \rho\rangle \bigvee_{i<\alpha} x_i = \bigvee_{i<\alpha} \langle\sigma, \rho\rangle x_i$, and:

**Trace closure**

(M) $\langle\sigma, \rho\rangle\langle\rho, \theta\rangle x \geq \langle\sigma, \theta\rangle x$ (S) $x \geq \langle\sigma, \sigma\rangle x$ (H) $\bigvee_{\sigma\in\mathbb{S}} \langle\sigma, \sigma\rangle x \geq x$

$$\frac{\forall\sigma.\, \xi_1^{?}\langle\sigma, \sigma\rangle\xi_2^{?}x \in K}{\xi_1^{?}\xi_2^{?}x \in K}$$

**Fig. 2.** The $\mathsf{hush}$ rule

The first two axiom schemes are algebraic counterparts to $\mathsf{mumble}$ and $\mathsf{stutter}$. These alone do not recover Brookes's model—the representation theorem for the theory without the H axiom includes potentially-empty traces. The axiom H fails in this model, but holds in Brookes's. In the representation theorem for $\mathsf{B}$ it is tempting to require, along with closure under Brookes's $\mathsf{mumble}$ and $\mathsf{stutter}$ trace deductions, closure under $\mathsf{hush}$: presented in fig. 2 for a set of traces $K$. However, there is no need, due to the non-emptiness of the traces. Indeed, either $\xi_1^{?}$ or $\xi_2^{?}$ must be non-empty for the rule to apply. Take $\sigma$ to match an adjacent transition, and apply the $\mathsf{mumble}$ closure rule to obtain the required consequence. This nuanced observation exposing the $\mathsf{hush}$ rule would be hard to notice without this algebraic analysis.

To conclude, we formulate the representation theorem for $\mathsf{B}$. Let $X \in \mathbf{Set}$. Define the $\Sigma_{\mathsf{B}}$-algebra $\mathbf{B}X$ with carrier $\underline{\mathbf{B}X} \coloneqq \mathcal{P}^{\dagger}(\mathsf{T}X)$ and interpretations:

$$\mathbf{B}X[\![\bigvee_{i<\alpha}]\!]_{\mathrm{op}} K_i \coloneqq \bigcup_{i<\alpha} K_i \qquad \mathbf{B}X[\![\langle\sigma, \rho\rangle]\!]_{\mathrm{op}} K \coloneqq \{\langle\sigma, \rho\rangle\tau \mid \tau \in K\}^{\dagger}$$

Additionally, define $\mathrm{return} : X \to \underline{\mathbf{B}X}$ by $\mathrm{return}\, x \coloneqq \lambda x.\, \{\langle\sigma, \sigma\rangle x \mid \sigma \in \mathbb{S}\}^{\dagger}$.

To prove that this is a free $\mathsf{B}$-model, we use reification as in §5.2, though here reification is more straightforward. A trace is reified as itself, and sets of

traces use countable-joins as before: $\mathrm{reify}\, K \coloneqq \left(X \vdash_{\Sigma_{\mathsf{B}}} \bigvee_{\tau \in K} \underline{\tau} : \star\right)$. The monad obtained from the next proposition is Brookes's model:

**Proposition 26.** *The pair* $\langle \mathbf{B}X, \mathrm{return}\rangle$ *is a free* $\mathsf{B}$*-model over* $X$*, for which the representation sends* $e : X \to \underline{\mathbf{A}}$ *to* $e^{\#} : \mathbf{B}X \to \mathbf{A}$ *by* $e^{\#}_{\square} K \coloneqq \mathbf{B}X[\![\mathrm{reify}_{\square}\, K]\!]_{\mathrm{term}} e$.

## 6.4 Translations and equivalences

We will need the following notions for relating presentations. Consider a map between two sort sets $\epsilon : \mathbf{sort}_1 \to \mathbf{sort}_2$. It lifts to $\epsilon : \mathbf{Set}^{\mathbf{sort}_2} \to \mathbf{Set}^{\mathbf{sort}_1}$ by precomposition: $(\epsilon \boldsymbol{Y})_{\square} \coloneqq \boldsymbol{Y}_{\epsilon\square}$. It forms the object part of a geometric morphism between (pre)sheaf toposes, i.e., it has left and right adjoints. The left adjoint $\epsilon^* : \mathbf{Set}^{\mathbf{sort}_1} \to \mathbf{Set}^{\mathbf{sort}_2}$ is in this case $(\epsilon^* \boldsymbol{X})_{\diamond} \coloneqq \coprod_{\epsilon\square = \diamond} \boldsymbol{X}_{\square}$. When $\epsilon$ is injective, the left adjoint is given by the simpler formula $\epsilon^* \boldsymbol{X} \coloneqq \{x : \epsilon\square \mid x \in \boldsymbol{X}_{\square}\}$.

*Example 27.* The geometric morphism for the map $\star \mapsto \circ : \{\star\} \rightarrowtail \{\bullet, \circ\}$ is the forgetful functor $(-)_{\circ} : \mathbf{Set}^{\{\bullet,\circ\}} \to \mathbf{Set}^{\{\star\}} \cong \mathbf{Set}$. As we saw in §6.2, its left adjoint is $(-)^{\circ} : \mathbf{Set}^{\{\star\}} \to \mathbf{Set}^{\{\bullet,\circ\}}$. □

Let $\Sigma_1$ and $\Sigma_2$ be signatures and $\epsilon : \mathbf{sort}_{\Sigma_1} \to \mathbf{sort}_{\Sigma_2}$ a map between their sort sets. A *translation of signatures* $\mathbf{E} : \Sigma_1 \rightarrowtail \Sigma_2$ *along* $\epsilon$ is an assignment, to each $(O : \square\langle \diamond_i \rangle_{i<\alpha}) \in \Sigma_1$, of a term $\mathbf{E}O \in \mathrm{Term}^{\Sigma_2}_{\epsilon\square} \{x_i : \epsilon\diamond_i \mid i < \alpha\}$. Such a translation yields a functor $\mathbf{E}_{\mathrm{tln}} : \mathbf{Alg}\Sigma_2 \to \mathbf{Alg}\Sigma_1$, mapping a $\Sigma_2$-algebra $\mathbf{B}$ to:

$$\underline{\mathbf{E}_{\mathrm{tln}}\mathbf{B}} \coloneqq \epsilon\underline{\mathbf{B}} \qquad \mathbf{E}_{\mathrm{tln}}\mathbf{B}\,[\![O : \square\langle \diamond_i \rangle_{i<\alpha}]\!]_{\mathrm{op}}\,\langle b_i \rangle \coloneqq \mathbf{B}\,[\![\mathbf{E}O]\!]_{\mathrm{term}}\,\langle x_i \mapsto b_i \rangle_{i<\alpha}$$

For a given family $\boldsymbol{Y} \in \mathbf{Set}^{\mathbf{sort}_{\Sigma_2}}$, such a translation therefore extends uniquely to a $\Sigma_1$-homomorphism $(\mathbf{E}_{\mathrm{tln}})_{\boldsymbol{Y}} : F_{\Sigma_1} \epsilon \boldsymbol{Y} \to \mathbf{E}_{\mathrm{tln}} F_{\Sigma_2} \boldsymbol{Y}$.

*Example 28.* We have a translation $\mathbf{E} : \Sigma_{\mathsf{G}} \rightarrowtail \Sigma_{\mathbb{S}}$ along $\star \mapsto \bullet : \{\star\} \rightarrowtail \{\bullet, \circ\}$ that translates the $\Sigma_{\mathsf{G}}$-operators using their respective copies in the $\bullet$ sort:

$$\begin{aligned}
\mathbf{E}(\textstyle\bigvee_{\alpha} : \alpha) &\coloneqq (\{x_i : \bullet \mid i < \alpha\} \vdash_{\Sigma_{\mathbb{S}}} \textstyle\bigvee_{i<\alpha} x_i &&: \bullet)\\
\mathbf{E}(\mathsf{L}_{\ell} : \mathbf{2}) &\coloneqq (\{x_0, x_1 : \bullet\} \vdash_{\Sigma_{\mathbb{S}}} \mathsf{L}_{\ell}(x_0, x_1) &&: \bullet)\\
\mathbf{E}(\mathsf{U}_{\ell,b} : \mathbf{1}) &\coloneqq (\{x_0 : \bullet\} \vdash_{\Sigma_{\mathbb{S}}} \mathsf{U}_{\ell,b}\, x_0 &&: \bullet)
\end{aligned}$$

□

A translation of *presentations* $\mathbf{E} : \mathfrak{p}_1 \rightarrowtail \mathfrak{p}_2$ along $\epsilon$ is a translation of their signatures along $\epsilon$ that, moreover, preserves the provability of axioms:

$$(\boldsymbol{X} \vdash_{\Sigma_{\mathfrak{p}_1}} t_1 = t_2 : \square) \in \mathrm{Ax}_{\mathfrak{p}_1} \implies \epsilon^* \boldsymbol{X} \vdash_{\mathfrak{p}_2} \mathbf{E}_{\mathrm{tln}} t_1 = \mathbf{E}_{\mathrm{tln}} t_2 : \epsilon\square$$

*Example 29.* The translation of global state into shared state from example 28 is a translation of presentations $\mathbf{E} : \mathsf{G} \rightarrowtail \mathbb{S}$. □

Translations along composable sort maps compose via substitution, and a translation $\mathbf{E} : \mathfrak{p} \rightarrowtail \mathfrak{p}$ along $\mathsf{id}_{\Sigma_{\mathfrak{p}}}$ is an *identity* translation when, for all terms $t \in \mathrm{Term}^{\Sigma_{\mathfrak{p}}}_{\square} \boldsymbol{X}$, we have $\boldsymbol{X} \vdash_{\mathfrak{p}} \mathbf{E}_{\mathrm{tln}} t = t : \square$. A translation $\mathbf{E} : \mathfrak{p}_1 \rightarrowtail \mathfrak{p}_2$ along $\epsilon$ is an *equivalence* if $\epsilon$ is a bijection, and there exists a translation $\mathbf{E}^{-1} : \mathfrak{p}_2 \rightarrowtail \mathfrak{p}_1$ along $\epsilon^{-1}$, such that $\mathbf{E} \circ \mathbf{E}^{-1}$ and $\mathbf{E}^{-1} \circ \mathbf{E}$ are identity translations. We then write $\mathfrak{p}_1 \simeq \mathfrak{p}_2$ and say that the presentations are *equivalent*. Two multi-sorted theories are equivalent iff their associated free-model monads are isomorphic.

### 6.5 Translation through the two-sorted theory of transitions

We define a single-sorted presentation $\mathsf{Tgs}$ of the *open* transitions $\{\!|\sigma, \rho|\!\}$ as sequential operators. The signature $\Sigma_{\mathsf{Tgs}}$ consists of countable-joins $\Sigma_{\mathsf{V}}$ and a unary open transition operator $(\!|\sigma, \rho|\!)$ for $\sigma, \rho \in \mathbb{S}$. The axioms $\mathrm{Ax}_{\mathsf{Tgs}}$ consist of the countable-join semilattice axioms $\mathrm{Ax}_{\mathsf{V}}$, strict distributivity axioms (ND-$\mathsf{T}$) $(\!|\sigma, \rho|\!) \bigvee_{i<\alpha} x_i = \bigvee_{i<\alpha} (\!|\sigma, \rho|\!)\, x_i$, and:

| Open transition axioms | | |
|---|---|---|
| | (Seq$^{=}$) $(\!|\sigma, \rho|\!)\, (\!|\rho, \theta|\!)\, x = (\!|\sigma, \theta|\!)\, x$ | |
| (HS) $x = \bigvee_{\sigma \in \mathbb{S}} (\!|\sigma, \sigma|\!)\, x$ | (Seq$^{\neq}$) $(\!|\sigma, \rho|\!)\, (\!|\mu, \theta|\!)\, x = \bot$ | $\rho \neq \mu$ |

Translate $\mathbf{E}_{\mathsf{G}} : \mathsf{Tgs} \rightarrowtail \mathsf{G}$ by interpreting transitions as the open transitions from §5.2: $\mathbf{E}_{\mathsf{G}}\, (\!|\sigma, \rho|\!) := (x_0 \vdash_{\Sigma_{\mathsf{G}}} \{\!|\sigma, \rho|\!\}\, x_0)$. Conversely, translate $\mathbf{E}_{\mathsf{Tgs}} : \mathsf{G} \rightarrowtail \mathsf{Tgs}$ as follows, similar to the representation of update and lookup from §5.2:

$$\mathbf{E}_{\mathsf{Tgs}} \mathsf{U}_{\ell,b} := (x_0 \vdash_{\Sigma_{\mathsf{Tgs}}} \textstyle\bigvee_{\sigma \in \mathbb{S}} (\!|\sigma, \sigma[\ell \mapsto b]|\!)\, x_0) \quad \mathbf{E}_{\mathsf{Tgs}} \mathsf{L}_{\ell} := (x_0, x_1 \vdash_{\Sigma_{\mathsf{Tgs}}} \textstyle\bigvee_{\sigma \in \mathbb{S}} (\!|\sigma, \sigma|\!)\, x_{\sigma_\ell})$$

Using the equivalence $\mathsf{Tgs} \simeq \mathsf{G}$ that these translations witness we can translate $\mathsf{B} \rightarrowtail \mathsf{S\!\!S}$ along $\star \mapsto \circ$. We define a two-sorted presentation $\mathsf{Tr}$, mimicking the definition of $\mathsf{S\!\!S}$ but replacing the operators and axioms of $\mathsf{G}$ with those of $\mathsf{Tgs}$ in the hold ($\bullet$) sort: $\mathrm{Ax}_{\mathsf{Tr}} := \boxed{\mathrm{Ax}^{\bullet}_{\mathsf{Tgs}}} \cup \mathrm{Ax}^{\circ}_{\mathsf{V}} \cup \{\text{ND-}\triangleright, \text{ND-}\triangleleft\} \cup \{\mathrm{Empty}, \mathrm{Fuse}\}$. Extending the translations $\mathbf{E}_{\mathsf{Tgs}}$ and $\mathbf{E}_{\mathsf{G}}$ to all of the operators gives an equivalence $\mathsf{Tr} \simeq \mathsf{S\!\!S}$. So $\mathsf{Tr}$ induces the same monad as $\mathsf{S\!\!S}$, recovering Brookes's model.

Define the translation $\mathbf{E}_{\mathsf{Tr}} : \mathsf{B} \rightarrowtail \mathsf{Tr}$ along $\star \mapsto \circ$ by sending transitions to their delimited open counterparts: $\mathbf{E}_{\mathsf{Tr}} \langle \sigma, \rho \rangle := (x_0 : \circ \vdash_{\Sigma_{\mathsf{Tr}}} \triangleleft\, (\!|\sigma, \rho|\!) \triangleright x_0 : \circ)$. Using $\mathsf{Tr} \simeq \mathsf{S\!\!S}$ we get $\mathsf{B} \rightarrowtail \mathsf{S\!\!S}$ (fig. 3). Brookes's model, as a free $\mathsf{B}$-model, is thus the $\circ$-sorted fragment of $\mathsf{S\!\!S}$ over $\circ$-variables, formally.

$$\begin{array}{ccccc} & & \mathsf{Tgs} & \simeq & \mathsf{G} \\ & & \downarrow & \overset{\star}{\downarrow}\bullet & \downarrow \\ \mathsf{B} & \overset{\star \mapsto \circ}{\rightarrowtail} & \mathsf{Tr} & \simeq & \mathsf{S\!\!S} \end{array}$$

**Fig. 3.** Th. chart

## 7 Conclusion and further work

We presented an equational theory for shared state ($\mathsf{S\!\!S}$). It separates reasoning into two layers. In the held layer ($\bullet$), we prohibit the concurrent environment from accessing memory, and we can reason about memory accesses by a pool of threads sequentially. In the ceded layer ($\circ$), the concurrent environment may interleave, and local memory access is forbidden. We also presented theories of transitions ($\mathsf{B}$, $\mathsf{Tgs}$, & $\mathsf{Tr}$) and formally related them to (non-deterministic) global state ($\mathsf{G}$) and shared state ($\mathsf{S\!\!S}$). The single-sorted theory $\mathsf{B}$ recovers Brookes's model, but it does so by using Brookes's `await` construct, which we find unnatural; and it does not admit global state explicitly as a component of the theory. We believe that admitting global state will inform modelling other effects in the concurrent setting. Our theory $\mathsf{S\!\!S}$ addresses these concerns. It admits the global state theory as-is, and axiomatises the mode-switching operators ($\triangleleft/\triangleright$) without explicit interaction with global state. This theory recovers Brookes's model exactly, in a principled manner: by transforming a monad and its operations along an adjunction; and, independently, through algebraic translations.

Our theory uses countable-join semilattices to recover Brookes's model. They can express iteration (i.e. `while`-loops). The same model admits first-order recursion, i.e. least-fixpoints of mutually-defined first-order functions, using the $\omega$-complete partial order structure of the refinement order and the Scott-continuity of the semantics. We can support higher-order recursion by recourse to domain-theory, generalising algebraic theories using order-enriched theories. There are several standard variants, each with subtle logical trade-offs [34]. We can also restrict the semantics to terminating languages by restricting to finite joins, and using finitely-generated closed subsets for the representation.

We want to analyse Brookes's parallel composition operator algebraically. Brookes composed programs in parallel by interleaving traces from each thread. Initial results show we can define Brookes's parallel composition by simultaneous induction over terms. However, we would like to provide a more abstract account, by recourse to the universal property of free models. This abstraction may expose special properties of global state, or lead to a general parallel composition operation satisfying the expected laws of concurrent programming [16, 31, 39].

We would like to model more effects within this modular multi-sorted algebraic framework. These effects include: more advanced notions of state, such as dynamic allocation [22], higher-order memory cells [28, 41], and weak memory [13, 14]; control-flow effects such as exceptions and effect handlers [4]; and probabilistic programming with shared state [26].

If the multi-sorted approach does indeed generalise to more sophisticated effects, then it will be instructive to review its assumptions. For example, the strictness axioms impose a partial-correctness discipline: the semantics says nothing about the effect a diverging program has on its memory. Relaxing or removing strictness may give a model that allows us to reason about diverging programs.

Our two sorts limit access to the whole store. We would like to explore finer granularity. For example, a theory with per-location access limitation, with sorts for every finite subset $s \subseteq \mathbb{L}$ of locations, and operators $(\lhd_\ell : s \backslash \{\ell\} \langle s \cup \{\ell\} \rangle)$ and $(\rhd_\ell : s \cup \{\ell\} \langle s \backslash \{\ell\} \rangle)$. We expect the axiomatisation's design to require subtlety.

It may be interesting to to expose the sort discipline in the surface language through typing judgements, explicating regions that rule out data-races with the environment. It seems such judgements would rule out deadlocks structurally, and so may limit expressiveness. Whether this idea is useful remains to be seen.

In conclusion, our two-sorted decomposition of Brookes's seminal model provides new insights into its assumptions and components, and reveals new directions for modelling more advanced features involving concurrent shared state.

**Acknowledgments.** Supported by the Israel Science Foundation (grant number 814/22) and the European Research Council (ERC) under the European Union's Horizon 2020 research and innovation programme (grant agreement no. 851811); and by a Royal Society University Research Fellowship and Enhancement Award. For the purpose of Open Access the authors have applied a CC BY public copyright licence to any Author Accepted Manuscript version arising from this submission. We thank Danel Ahman, Andrej Bauer, Martín Escardó, Justus Matthiesen, Sam Staton, and Rob van Glabbeek for interesting and useful discussions and suggestions.

## A No-go results

We can present Brookes's model using a single-sorted presentation (§6.3). However, we found this presentation unsatisfactory, and so propose a two-sorted account. Our use of the two-sorted approach follows a relatively thorough investigation into alternative single-sorted approaches, and we can provide some crisp results that certain single-sorted approaches fail. These no-go results, together with the perspectives on future work the two-sorted decomposition suggests (§7), are evidence for the merit of our two-sorted approach. They may also inform future search for a single-sorted presentation that we have overlooked.

Single-sorted transitions present Brookes's model in terms of the `await` construct. This presentation highlights the importance of `await` for reasoning in Brookes's model and why `await` is a key ingredient in Brookes's full abstraction result. Without `await`, Brookes's model is not fully abstract at $1^{\text{st}}$-order:

**No-go 1 (Svyatlovskiy et al. [42]).** *Brookes's model is not fully-abstract w.r.t. the operational semantics in which differentiating contexts can only read and mutate single memory cells atomically.*

Moreover, every single-sorted presentation of Brookes's model must involve operators other than the interpretations of read and write, considered as generic effects [36]. Formally, given a family of algebraic operations and a monad, we can construct the sub-monad generated by a set of operations [21, 23, 24].

**No-go 2.** *The sub-monad generated by the semantics of read and write, and by union, differs from the Brookes model.*

*Proof.* The trace-sets generated by read and write always contain a trace in which at most one cell changes within each transition. Brookes's model includes other subsets, definable via the `await` construct. □

The traces in Brookes's model explicitly yield control to their concurrent environment. Following Abadi and Plotkin [1], we investigated adding an additional unary operator $\mathsf{Y}$ for yielding control to the concurrent environment. It is natural to interpret $\mathsf{Y}$ as adding a no-op transition $\langle\sigma,\sigma\rangle$ before every trace in its argument, modelling a possible interference by the environment. An alternative choice is to add such no-op transitions and also keep the original traces, modelling a *possibility* for a yield in the previous sense. Both of these options trivialise in Brookes's model:

**No-go 3.** *Consider the following interpretations of* $\mathsf{Y}$ *in Brookes's model:*

$$[\![\mathsf{Y}]\!]^{1}_{\mathrm{op}}\, K \coloneqq \{\langle\sigma,\sigma\rangle\tau \mid \tau \in K\} \qquad [\![\mathsf{Y}]\!]^{2}_{\mathrm{op}}\, K \coloneqq K \cup [\![\mathsf{Y}]\!]^{1}_{\mathrm{op}}\, K$$

*Then* $[\![\mathsf{Y}]\!]^{i}_{\mathrm{op}}\, K = K$ *for both* $i \in \{1,2\}$*, for any closed* $K$*.*

*Proof.* $K$ is closed under $\mathsf{stutter}$ and $\mathsf{hush}$. □

Even though Brookes's model does not support this intuition, we explored where the yield approach leads. With this yield operator, lookup and update can represent interference-free memory-access as axiomatised in the global-state theory, and surface-language level read and write can be modelled by some combination of the algebraic operators. Formally, let $\mathsf{Res}$ be a presentation that includes non-deterministic global state, and the yield operator $\mathsf{Y}$, which is $\mathsf{Res}$-provably strict and distributes over joins.

**Option 1 (Dvir et al.'s presentation [12]).** For a previous theory of ours, we took a minimal $\mathsf{Res}$ satisfying our restrictions, and defined the algebraic representation of read:

$$\mathsf{R}_\ell(x_0, x_1) \coloneqq \left(x_0, x_1 \vdash_{\Sigma_{\mathsf{Res}}} \mathsf{L}_\ell((x_0 \vee \mathsf{Y}\, x_0), (x_1 \vee \mathsf{Y}\, x_1))\right)$$

Reading *may* admit an interference point after looking the value up in memory.

**Option 2 (Plotkin's presentation [33]).** Another natural option is to take $\mathsf{Res}$ to also prove that $\mathsf{Y}$ is a closure operator, i.e. $x \vdash_{\mathsf{Res}} \mathsf{Y}\,\mathsf{Y}\, x = \mathsf{Y}\, x \geq x$. In this option, the intuition is that $\mathsf{Y}$ *potentially* yields, and yielding successively is immaterial. This theory allows the algebraic representation of read to be a bit more natural:

$$\mathsf{R}_\ell(x_0, x_1) \coloneqq \left(x_0, x_1 \vdash_{\Sigma_{\mathsf{Res}}} \mathsf{Y}\, \mathsf{L}_\ell(\mathsf{Y}\, x_0, \mathsf{Y}\, x_1)\right)$$

Both options prove (Irrelevant Read Elim), but not (Irrelevant Read Intro):

$$x \vdash_{\mathsf{Res}} \mathsf{R}_\ell(x, x) \geq x \qquad \text{(Irrelevant Read Elim)}$$

$$x \not\vdash_{\mathsf{Res}} \mathsf{R}_\ell(x, x) \leq x \qquad \text{(Irrelevant Read Intro)}$$

Brookes's model validates (Irrelevant Read Intro), so the proposed theories are both not abstract enough. Adding (Irrelevant Read Intro) as an axiom in either version is problematic, as it implies the following inequation:

$$x_{0,0}, x_{0,1}, x_{1,0}, x_{1,1} \vdash_{\Sigma_{\mathsf{Res}}} \mathsf{R}_\ell(\mathsf{R}_\ell(x_{0,0}, x_{0,1}), \mathsf{R}_\ell(x_{1,0}, x_{1,1})) \leq \mathsf{R}_\ell(x_{0,0}, x_{1,1}) \qquad \text{(Same Read Intro)}$$

The corresponding program transformation is invalid in our setting because the environment can interfere, mutating $\ell$ between the successive reads.

We summarise this intermediate result:

**No-go 4.** *Let* $\mathsf{Res}$ *be either Dvir et al.'s or Plotkin's presentation, and define* $\mathsf{R}_\ell$ *accordingly. If* (Irrelevant Read Elim) *and* (Irrelevant Read Intro) *are valid in* $\mathsf{Res}$*, then so is* (Same Read Intro)*.*

Another approach is to add unary operators $\lhd'$ and $\rhd'$ that delimit the memory accesses. Formally, let $\mathsf{Del}$ be a presentation that includes non-deterministic global state, and the delimiting operators $\lhd'$ and $\rhd'$, which are $\mathsf{Del}$-provably strict and distribute over joins. Define the algebraic representation of read:

$$\mathsf{R}_\ell(x_0, x_1) \coloneqq \left(x_0, x_1 \vdash_{\Sigma_{\mathsf{Del}}} \lhd'\, \mathsf{L}_\ell(\rhd'\, x_0, \rhd'\, x_1)\right) \qquad (\star)$$

This approach subsumes the two $\mathsf{Res}$ options suggested above, by using the axioms $x \vdash \lhd'\, x = x$ and $x \vdash \rhd'\, x = x \vee \mathsf{Y}\, x$ for Dvir et al.'s presentations; and using $x \vdash \lhd'\, x = \mathsf{Y}\, x$ and $x \vdash \rhd'\, x = \mathsf{Y}\, x$ for Plotkin's presentation. In both cases, and more generally whenever $\lhd'$ and $\rhd'$ are given by a combination of joins and yields, they commute:

**Lemma 30.** *Let $t_1$ and $t_2$ be $\{\vee, \mathsf{Y}\}$-terms over $\{x\}$. If $x \vdash_{\mathsf{Del}} \lhd'\, x = t_1$ and $x \vdash_{\mathsf{Del}} \rhd'\, x = t_2$, then $x \vdash_{\mathsf{Del}} \lhd'\, \rhd'\, x = \rhd'\, \lhd'\, x$.*

*Proof.* Using the semilattice axioms and distributivity of $\mathsf{Y}$ over joins, every $\{\vee, \mathsf{Y}\}$-term $t$ over $\{x\}$ is $\mathsf{Del}$-equal to a non-deterministic choice between terms of the form $\mathsf{Y}^n\, x$ for $n \in N_t \subseteq \mathbb{N}$. Both terms above are equal to the same term of this form, with $N = \{n_1 + n_2 \mid n_1 \in N_{\lhd'\, x}, n_2 \in N_{\rhd'\, x}\}$. ☐

Any alternative of $\mathsf{Del}$ for which $\lhd'$ and $\rhd'$ commute is not satisfactory:

**No-go 5.** *Let $\mathsf{Del}$ be a presentation that includes non-deterministic global state, and the unary operators $\lhd'$ and $\rhd'$, which $\mathsf{Del}$ proves to be strict, distribute over joins, and commute. With read from ($\star$), if $\mathsf{Del}$ proves (Irrelevant Read Elim) and (Irrelevant Read Intro), then it proves (Same Read Intro).*

*Proof.* Combining (Irrelevant Read Elim) and (Irrelevant Read Intro), we have $x \vdash_{\mathsf{Del}} \mathsf{R}_\ell(x, x) = x$. Using global-state, we have $x \vdash_{\mathsf{Del}} \mathsf{R}_\ell(x, x) = \lhd'\, \rhd'\, x$. Therefore, $x \vdash_{\mathsf{Del}} \lhd'\, \rhd'\, x = x$. They commute, so $x \vdash_{\mathsf{Del}} \rhd'\, \lhd'\, x = x$. Using global-state, we prove (Same Read Intro) in $\mathsf{Del}$. ☐

Therefore, any such theory $\mathsf{Del}$ is either unsound, or it fails to validate a transformation that Brookes's model does. Thus, when picking $\mathsf{Del}$, we need to make sure that $\lhd'$ and $\rhd'$ do not commute.

As a final option we cover here, we could take the axioms $x \vdash \lhd'\, \rhd'\, x = x$ and $x \vdash \rhd'\, \lhd'\, x \geq x$. These are like the closure pair axioms of our shared-state presentation $\mathbb{S}$, but without the sort discipline. The single-sorted signature allows ill-bracketed terms such as $x \vdash \lhd'\, \lhd'\, x$. Though it may be possible to axiomatise that all such terms are equal to $\bot$, a more principled way to avoid such terms is to use a two-sorted theory as we have.

The analysis we offered in this section does not rule out the possibility of a satisfactory single-sorted theory of shared-state. We hope that these considerations could inform future pursuit of this theory, or a tighter no-go result.

## B Distributivity

This section is devoted to the technical definition of distributivity.

Let $\Sigma$ be a multi-sorted signature, $(P : \boxbox \langle \diamonddiamond_i \rangle_{i<\alpha}) \in \Sigma$ be an operator, and $i_0 < \alpha$ be one of the positions in $P$'s scheme. Assume further that both $\diamonddiamond_{i_0}$ and $\boxbox$ have 'single-sorted' operators $(S : \diamonddiamond_{i_0}(\beta \cdot \diamonddiamond_{i_0})), (S' : \boxbox(\beta \cdot \boxbox)) \in \Sigma$ with the same arity length $\beta$. We define the following *distributivity* axiom [18]:

$$\{x_i : \Diamond_i \mid i_0 \neq i < \alpha\} \cup \{y_j : \Diamond_{i_0} \mid j < \beta\} \vdash_\Sigma$$
$$P \left\langle \begin{cases} i \neq i_0 : & x_i \\ i = i_0 : & S \left\langle y_j \right\rangle_j \end{cases} \right\rangle_i = S' \left\langle P \left\langle \begin{cases} i \neq i_0 : & x_i \\ i = i_0 : & y_j \end{cases} \right\rangle_i \right\rangle_j : \Box$$

which we call the *distributivity of* $P$ *over* $S, S'$ *in the* $i_0$*-component.*

Distributivity over binary joins implies monotonicity, in the following sense. Let $\mathfrak{p}$ be a presentation, $(O : \Box \left\langle \Diamond_i \right\rangle_{i<\alpha}) \in \Sigma_{\mathfrak{p}}$ be an operator, and $i_0 < \alpha$ an index into its sorting scheme. Assume $\Box, \Diamond_{i_0}$ include the theory of semilattices, and that $O$ distributes over the binary joins of $\Diamond_{i_0}$ and $\Box$ in the $i_0^{\text{th}}$ component. Then $O$ is monotone in this component w.r.t. the semilattice preorder, i.e., the following deduction rule is admissible:

$$\frac{\boldsymbol{Y} \vdash_{\mathfrak{p}} l \leq r : \Diamond_{i_0}}{\{x_i : \Diamond_i \mid i_0 \neq i < \alpha\} \cup \boldsymbol{Y} \vdash_{\mathfrak{p}} O \left\langle \begin{cases} i \neq i_0 : & x_i \\ i = i_0 : & l \end{cases} \right\rangle_i \leq O \left\langle \begin{cases} i \neq i_0 : & x_i \\ i = i_0 : & r \end{cases} \right\rangle_i}$$

Specifically, if $\mathfrak{p}$ includes the theory of semilattices in all sorts, and every operator distributes over binary joins, then the congruence rule for inequations is valid.

## C Proof of the representation theorem

To start, we first prove proposition 23, soundness of encoded trace deductions:

*Proof.* First, standardly in $\mathsf{G}$ we have $x : \star \vdash_{\mathsf{G}} \{\!\{\sigma, \rho\}\!\} \{\!\{\rho', \theta\}\!\}\, x \geq \{\!\{\sigma, \theta\}\!\}\, x : \star$ and $x : \star \vdash_{\mathsf{G}} \{\!\{\sigma, \sigma\}\!\}\, x \geq x : \star$, which are included in the $\bullet$ sort in $\mathbb{S}$.

– The former, combined with Fuse, leads to soundness of $\mathsf{mumble}$.
– The latter, combined with Empty, leads to soundness of $\mathsf{stutter}$. □

That reification is indifferent to closure follows:

**Proposition 31.** *For* $K \in \mathcal{P}^{\aleph_0}_{\Box}(\mathbb{T}\boldsymbol{X})$, $\boldsymbol{X} \vdash_{\mathbb{S}} \mathrm{reify}_{\Box} K = \mathrm{reify}_{\Box} K^{\dagger} : \Box$.

*Proof.* Follows from proposition 23 by inequational reasoning. □

To prove the $\mathbb{S}$-Rep. Thm., let $\boldsymbol{X} \in \mathbf{Set}^{\{\bullet,\circ\}}$. We start by giving alternative formulas to the interpretations of the lock operators.

**Lemma 32.** *Denote the set of sequences of transitions, where each transition has equal components* $\mathbb{S}^*_{=} \coloneqq \{\langle\sigma, \sigma\rangle \mid \sigma \in \mathbb{S}\}^*$. *The following hold:*

$$\mathbf{R}\boldsymbol{X} [\![\lhd]\!]_{\mathrm{op}} K = \left\{ \circ\xi_0^{?}\xi \Diamond x \;\middle|\; \xi_0^{?} \in \mathbb{S}^*_{=}, \bullet\xi\Diamond x \in K \right\}$$
$$\mathbf{R}\boldsymbol{X} [\![\rhd]\!]_{\mathrm{op}} K = \{ \bullet\xi\Diamond x, \bullet\langle\sigma, \sigma\rangle\xi\Diamond x \mid \sigma \in \mathbb{S}, \circ\xi\Diamond x \in K \}$$

*Proof sketch.* The fact that $K$ is closed means that most trace deductions afforded in the interpretations as defined in the $\mathbb{S}$-Rep. Thm. are redundant.

– In $\mathbf{R}\boldsymbol{X}\,[\![\lhd]\!]_{\mathrm{op}}\,K$, the only application of a trace deduction that results in a trace that would is not in the set before taking the closure is one of `stutter` at the start of the trace.
– In $\mathbf{R}\boldsymbol{X}\,[\![\rhd]\!]_{\mathrm{op}}\,K$, the only application of a trace deduction that results in a trace that would is not in the set before taking the closure is one of `mumble` at the start of the trace. ☐

**Lemma 33.** $\mathbf{R}\boldsymbol{X}$ *is an* $\boldsymbol{\mathbb{S}}$*-model.*

*Proof.* This amounts to showing that $\mathbf{R}\boldsymbol{X}$ validates every $\boldsymbol{\mathbb{S}}$-axiom.

– The countable-join semilattice ones follow standardly for sets and unions.
– Commutativity follows from the fact that interpretations are all defined by direct images.
– The global state equations validate as they did in the model from Dvir et al. [12], where they were interpreted in a similar manner.

This leaves Empty:

$$\begin{aligned}[\![\lhd]\!]\ [\![\rhd]\!]\ K &= [\![\lhd]\!]\,\{\bullet\xi\diamond x, \bullet\langle\sigma,\sigma\rangle\xi\diamond x \mid \sigma\in\mathbb{S}, \circ\xi\diamond x\in K\} \\ &= \left\{\circ\xi_0^?\xi\diamond x \;\middle|\; \xi_0^?\in\mathbb{S}^*_=, \bullet\xi\diamond x\in K\right\} = K\end{aligned}$$

where the last step is due to $K$ being closed; and Fuse:

$$\begin{aligned}[\![\rhd]\!]\ [\![\lhd]\!]\ K &= [\![\rhd]\!]\left\{\circ\xi_0^?\xi\diamond x \;\middle|\; \xi_0^?\in\mathbb{S}^*_=, \bullet\xi\diamond x\in K\right\} \\ &= \left\{\bullet\xi_0^?\xi\diamond x, \bullet\langle\sigma,\sigma\rangle\xi_0^?\xi\diamond x \;\middle|\; \sigma\in\mathbb{S}, \xi_0^?\in\mathbb{S}^*_=, \bullet\xi\diamond x\in K\right\} \supseteq K\end{aligned}$$

where the last step is by taking an empty $\xi_0^?$ in the first element. ☐

We mention some equations regarding open transitions provable in $\boldsymbol{\mathbb{S}}$.

**Lemma 34.** $x : \bullet \vdash_{\boldsymbol{\mathbb{S}}} \bigvee_{\sigma\in\mathbb{S}} \{\!|\sigma,\sigma|\!\}\, x = x : \bullet$

*Proof.* Follows from the global state validity: $x : \star \vdash_{\mathsf{G}} \bigvee_{\sigma\in\mathbb{S}} \{\!|\sigma,\sigma|\!\}\, x = x : \star$. ☐

**Lemma 35.** $x : \circ \vdash_{\boldsymbol{\mathbb{S}}} \bigvee_{\sigma\in\mathbb{S}} \lhd \{\!|\sigma,\sigma|\!\} \rhd x = x : \circ$

*Proof.* Follows from ND-$\lhd$, lemma 34, and Empty. ☐

Let's turn to the extension of environments along return. Let $\mathbf{A}$ be an $\boldsymbol{\mathbb{S}}$-algebra, and let $e : \boldsymbol{X} \to \underline{\mathbf{A}}$ be an $\boldsymbol{X}$-environment in $\mathbf{A}$. Then:

**Lemma 36.** $e^{\#}$ *is homomorphic.*

*Proof.* By evaluating both sides, it suffices to show that for every operator $(O : \boxdot\langle\boxdot_1, \dots, \boxdot_\alpha\rangle) \in \Sigma_{\boldsymbol{\mathbb{S}}}$, and all $K_i \in \underline{\mathbf{R}\boldsymbol{X}}_{\boxdot_i}$:

$$\boldsymbol{X} \vdash_{\boldsymbol{\mathbb{S}}} \mathrm{reify}(\mathbf{R}\boldsymbol{X}\,[\![O]\!]_{\mathrm{op}}\,(K_1, \dots, K_\alpha)) = O(\mathrm{reify}\,K_1, \dots, \mathrm{reify}\,K_\alpha) : \boxdot$$

As in the proof of lemma 33, most follow as in Dvir et al.'s model [12], and we focus again on the interesting cases of $\lhd$ and $\rhd$. In both cases, we use proposition 31 to simplify. For the treatment of the $\rhd$ case below, we use lemma 34 in the third equation:

$$\begin{aligned}
\boldsymbol{X} \vdash_{\mathbb{S}} \mathrm{reify}(\mathbf{R}\boldsymbol{X}\,[\![\rhd]\!]_{\mathrm{op}}\,K) &= \mathrm{reify}\,\{\bullet\langle\sigma,\sigma\rangle\xi\Diamond x \mid \sigma\in\mathbb{S}, \circ\xi\Diamond x\in K\} \\
&= \textstyle\bigvee_{\sigma\in\mathbb{S},\circ\xi\Diamond x\in K} \{\!\!\{\sigma,\sigma\}\!\!\} \rhd \underline{\circ\xi\Diamond x} \\
&= \textstyle\bigvee_{\circ\xi\Diamond x\in K} \rhd \underline{\circ\xi\Diamond x} \\
&= \rhd \textstyle\bigvee_{\circ\xi\Diamond x\in K} \underline{\circ\xi\Diamond x} = \rhd(\mathrm{reify}\,K) : \bullet \\
\boldsymbol{X} \vdash_{\mathbb{S}} \mathrm{reify}(\mathbf{R}\boldsymbol{X}\,[\![\lhd]\!]_{\mathrm{op}}\,K) &= \mathrm{reify}\,\{\circ\xi\Diamond x \mid \bullet\xi\Diamond x\in K\} \\
&= \textstyle\bigvee_{\bullet\xi\Diamond x\in K} \lhd \underline{\bullet\xi\Diamond x} \\
&= \lhd \textstyle\bigvee_{\bullet\xi\Diamond x\in K} \underline{\bullet\xi\Diamond x} = \lhd(\mathrm{reify}\,K) : \circ \qquad \square
\end{aligned}$$

**Lemma 37.** $e = e^{\#} \circ \mathrm{return}$ *for all* $x \in \boldsymbol{X}$.

*Proof.* By evaluating in $e$ the equations $x : \boxdot \vdash_{\mathbb{S}} \mathrm{reify}_{\boxdot}(\mathrm{return}_{\boxdot}\, x) = x : \boxdot$, which are easily verified in light of proposition 31, using lemmas 34 and 35. $\square$

**Lemma 38.** $\mathrm{return}^{\#} : \mathbf{R}\boldsymbol{X} \to \mathbf{R}\boldsymbol{X}$ *is the identity.*

*Proof sketch.* Follows by calculation, mainly by showing that for any $K \in \underline{\mathbf{R}\boldsymbol{X}}_{\bullet}$, we have that $\mathbf{R}\,\{x : \bullet\}\,[\![\{\!\!\{\sigma,\rho\}\!\!\}\, x]\!]_{\mathrm{term}}\,(x \mapsto K) = (\sigma,\rho)\,K$. $\square$

Finally, we show uniqueness. Let $f : \mathbf{R}\boldsymbol{X} \to \mathbf{A}$ be a homomorphism. Then:

**Lemma 39.** *If* $e = f \circ \mathrm{return}$ *then* $f = e^{\#}$.

*Proof.* We use the following notation. For any $\mathbb{S}$-algebra $\mathbf{B}$ and $\tilde{e} : \boldsymbol{X} \to \underline{\mathbf{B}}$, we denote $\mathrm{eval}(\tilde{e}) \coloneqq \mathbf{B}[\![-]\!]_{\mathrm{term}}\tilde{e} : \mathbf{F}^{\Sigma_{\mathbb{S}}}\boldsymbol{X} \to \mathbf{B}$. Thus, $\tilde{e}^{\#} = \mathrm{eval}(\tilde{e}) \circ \mathrm{reify}$.

Since $\mathrm{eval}(f \circ \mathrm{return}) : \mathbf{F}^{\Sigma_{\mathbb{S}}}\boldsymbol{X} \to \mathbf{A}$ is the only homomorphic extension of $f \circ \mathrm{return} : \boldsymbol{X} \to \underline{\mathbf{A}}$ along the inclusion $\iota : \boldsymbol{X} \hookrightarrow \mathrm{Term}^{\Sigma_{\mathbb{S}}}\boldsymbol{X}$, we have that $\mathrm{eval}(f \circ \mathrm{return}) = f \circ \mathrm{eval}(\mathrm{return})$. Using lemma 38:

$$e^{\#} = \mathrm{eval}(e) \circ \mathrm{reify} = \mathrm{eval}(f \circ \mathrm{return}) \circ \mathrm{reify} = f \circ \mathrm{eval}(\mathrm{return}) \circ \mathrm{reify} = f \qquad \square$$

Putting everything together, $\langle \mathbf{R}\boldsymbol{X}, \mathrm{return}\rangle$ is a $\mathbb{S}$-model over $\boldsymbol{X}$ (lemma 33) such that every environment homomorphically (lemma 36) extends along return (lemma 37), and does so uniquely (lemma 39). So $\langle \mathbf{R}\boldsymbol{X}, \mathrm{return}\rangle$ is a *free* $\mathbb{S}$-model over $\boldsymbol{X}$, proving the $\mathbb{S}$-Rep. Thm.